\documentclass[aps,showpacs,nofootinbib]{revtex4}
\usepackage{graphicx}
\usepackage{dcolumn}
\usepackage{bm}
\begin{document}
\newcommand{\newc}{\newcommand}
\newc{\ra}{\rightarrow}
\newc{\lra}{\leftrightarrow}
\newc{\beq}{\begin{equation}}
\newc{\eeq}{\end{equation}}
\newc{\barr}{\begin{eqnarray}}
\newc{\earr}{\end{eqnarray}}
\newcommand{\Od}{{\cal O}}
\newcommand{\lsim}   {\mathrel{\mathop{\kern 0pt \rlap
  {\raise.2ex\hbox{$<$}}}
  \lower.9ex\hbox{\kern-.190em $\sim$}}}
\newcommand{\gsim}   {\mathrel{\mathop{\kern 0pt \rlap
  {\raise.2ex\hbox{$>$}}}
  \lower.9ex\hbox{\kern-.190em $\sim$}}}
  \def\rpm{R_p \hspace{-0.8em}/\;\:}
\title {DIRECT DARK MATTER EVENT RATES WITH A VELOCITY DISTRIBUTION IN THE EDDINGTON APPROACH}
\author{ J.D. Vergados$^{1}$\footnote{Corresponding author} and D. Owen$^{2}$}
\affiliation{$^1$University of Ioannina, Ioannina, GR 45110, Greece \footnote {e-mail: vergados@cc.uoi.gr}.\\
 $^2$ Department of Physics, Ben Gurion University, Israel \footnote {e-mail: daowen137@gmail.com}.
}
\vspace{0.5cm}
 \begin{abstract}
Exotic dark matter together with the vacuum energy (associated
with the cosmological constant) seem to dominate the Universe.
Thus its direct detection is central to particle physics and
cosmology. Supersymmetry provides a natural dark matter candidate,
the lightest supersymmetric particle (LSP). One essential
ingredient in obtaining the direct detection rates is the density
and the velocity distribution of the LSP in our vicinity. In the
present paper we study simultaneously density profiles and
velocity distributions in the context of  the Eddington approach.
In such an approach, unlike the commonly assumed Maxwell-Boltzmann
(M-B) distribution, the upper bound of the velocity arises
naturally from the potential.
\end{abstract}
\maketitle
\section{Introduction}
The combined MAXIMA-1 \cite{MAXIMA-1}, BOOMERANG \cite{BOOMERANG},
DASI \cite{DASI} and COBE/DMR Cosmic Microwave Background (CMB)
observations \cite{COBE} imply that the Universe is flat
\cite{flat01}, $\Omega=1.11\pm0.07$ and that most of the matter in
the Universe is Dark \cite{SPERGEL}. i.e. exotic. Combining the
WMAP data with other experiments, crudely speaking one finds:
$$\Omega_b=0.05, \Omega _{CDM}= 0.30, \Omega_{\Lambda}= 0.65$$
Since the non exotic component cannot exceed $40\%$ of the CDM
~\cite {Benne}, there is room for the exotic WIMP's (Weakly
Interacting Massive Particles).
 Supersymmetry naturally provides candidates for the dark matter constituents
 \cite{GOODWIT}-\cite{ELLROSZ}.
 In the most favored scenario of supersymmetry the
LSP (Lightest Supersymmetric Particle) can be simply described as a Majorana fermion, a linear
combination of the neutral components of the gauginos and
higgsinos \cite{GOODWIT}-\cite{ref2}. In most calculations the
neutralino is assumed to be primarily a gaugino, usually a bino.
 Even though there exists firm indirect evidence for a halo of dark matter
 in galaxies from the
 observed rotational curves, it is essential to directly
detect \cite{GOODWIT}-\cite{KVprd}
 such matter.
Until dark matter is actually detected, we shall not be able to
exclude the possibility that the rotation curves result from a
modification of the laws of nature as we currently view them.  This makes it imperative that we
invest a
maximum effort in attempting to detect dark matter whenever it is
possible. Furthermore such a direct detection will also
 unravel the nature of the constituents of dark matter.\\
 The
 possibility of such detection, however, depends on the nature of the dark matter
 constituents (WIMPs).
  Since the WIMP is expected to be very massive, $m_{\chi} \geq 30 GeV$, and
extremely non relativistic with average kinetic energy $T \approx
50KeV (m_{\chi}/ 100 GeV)$, it can be directly detected
~\cite{GOODWIT}-\cite{KVprd} mainly via the recoiling of a nucleus
(A,Z) in elastic scattering. The event rate for such a process can
be computed from the following ingredients:
\begin{enumerate}
\item An effective Lagrangian at the elementary particle (quark)
level obtained in the framework of supersymmetry as described ,
e.g., in Refs~\cite{ref2,JDV96}.
\item A well defined procedure for transforming the amplitude
obtained using the previous effective Lagrangian from the quark to
the nucleon level, i.e. a quark model for the nucleon. This step
is not trivial, since the obtained results depend crucially on the
content of the nucleon in quarks other than u and d. This is
particularly true for the scalar couplings, which are proportional
to the quark masses~\cite{Dree}$-$\cite{Chen}, \cite{JDV06} as well as the
isoscalar axial coupling \cite{JELLIS93,JDV06}.
\item Knowledge of the relevant nuclear matrix elements
\cite{Ress}$-$\cite{DIVA00}, obtained with as reliable as possible many
body nuclear wave functions. Fortunately in the case of the scalar
coupling, which is viewed as the most important, the situation is
a bit simpler, since  then one  needs only the nuclear form
factor.
\item Knowledge of the WIMP density in our vicinity and its velocity distribution. Since the
essential input here comes from the rotational curves,  dark matter candidates other than the
LSP (neutralino) are also characterized
by similar parameters.
\end{enumerate}

In the past various velocity distributions have been considered.
The one most used is the isothermal Maxwell-Boltzmann velocity
distribution with $<\upsilon ^2>=(3/2)\upsilon_0^2$ where
$\upsilon_0$ is the velocity of the sun around the galaxy, i.e.
$220~km/s$. Extensions of this M-B distribution were also
considered, in particular those that were axially symmetric with
enhanced dispersion in the galactocentric direction
 \cite {Druk,Verg00}. In such
distributions an upper cutoff $\upsilon_{esc}=2.84\upsilon_0$ was introduced
by hand.

Non isothermal models have also been considered. Among those one should
mention the late infall of dark matter into the galaxy, i.e caustic rings
 \cite{SIKIVI1,SIKIVI2,Verg01,Green,Gelmini}, dark matter orbiting the
 Sun \cite{Krauss}, Sagittarius dark matter \cite{GREEN02}.

The correct approach in our view is to consider the Eddington
proposal \cite{EDDIN}, i.e. to obtain both the density and the
velocity distribution from a mass distribution, which depends both
on the velocity and the gravitational potential. Our motivation in
using Eddington \cite{EDDIN} approach to describing the density of
dark matter is found, of course, in his success in describing the
density of stars in globular clusters. Since this approach
adequately describes the distribution of stars in a globular
cluster in which the main interaction is gravitational and because
of its generality , we see no reason why such an approach should
not be applicable to dark matter that also interact
gravitationally. It should be noted that the attempt to use
Maxwellian (M-B) distribution to describe the star distribution in
globular clusters led to results that did not correspond to
observations \cite{EDDIN}.  So it seems that the use of M-B
distribution to describe dark matter is not very well motivated
and a different approach is required.

It seems, therefore, not surprising that this  approach has been
used by Merritt \cite{MERRITT} and  applied to dark matter by
Ullio and Kamionkowski\cite{ULLIO} and more recently by us
\cite{OWVER}. It is the purpose of the present paper to obtain a
dark matter velocity distribution, which is consistent with
assumed halo matter distributions and has a natural upper velocity
cut off.
The results presented here are motivated by the dark matter
candidate provided by supersymmetry, namely the LSP (neutralino). They can easily be extended, however,
 to be applied to other heavy WIMP candidates.
\section{The Dark Matter Distribution in the Context of  the Eddington approach}
As we have seen in the introduction the matter distribution can be given
 as follows
\beq dM=2\pi~f(\Phi({\bf r}),\upsilon_r,\upsilon_t)~dx~dy~dz
~\upsilon_t~d\upsilon_t~d\upsilon_r \label{distr.1} \eeq where the
function $f$ the distribution function, which depends on ${\bf r}$
through the potential $\Phi({\bf r})$ and the tangential and
radial velocities $\upsilon_t$ and $\upsilon_r$. In general the
distribution function is not symmetric. In the above expression we
assumed that it is only axially symmetric, with the two tangential
components being equal. Thus the density of matter $\rho$
satisfies the equation:
  \beq
d\rho=2\pi~f(\Phi({\bf r}
),\upsilon_r,\upsilon_t)~~\upsilon_t~d\upsilon_t~d\upsilon_r
\label{distr.2} \eeq
 It is more convenient instead of the
velocities to use the total energy $E$ and the angular momentum
$J$ via the equations \beq
J=\upsilon_t~r~~,~~2E=\upsilon_r^2+\frac{J^2}{r^2}+2~\Phi(r)
\label{distr.3} \eeq The use of these variables, which are
constants of motion, is very useful, when one wants to study
steady states. In doing this Eddington used  the result of Jeans
\cite{JEANS} that the density must be a function of first
integrals of the equations of motion which follows from
Liouville's theorem.  The advantage of this approach is that the
density can be 'inverted' and the velocity distribution can be
found. Following this approach we  find \beq
\rho=\frac{2\pi}{r^2}~\int~\int \frac{f(E,J)~J}{\sqrt{2(E-\Phi(r))-J^2/r^2}}
                                  ~dJ~dE
\label{distr.5}
\eeq
The limits of integration for $E$ are from $\Phi$ to $0$ and for $J$ from
$0$ to $[2r^2(E-\Phi(r))]^{1/2}$. \\
 Furthermore if the distribution function is known one can obtain
 the velocity distribution at some point, e.g. in our vicinity, by
 $$f(\Phi({\bf r}),\upsilon_r,\upsilon_t)|_{{\bf r=r_s}}$$
 The problem which is more interesting is: Can one obtain the distribution function
 given the density (and hence the potential via Poisson's equation)?
 The answer is affirmative via Eddington's treatment of the distribution
 function and quite easy,
  if the distribution does not explicitly depend on J, but is only a function of E.
 In the present work we will be concerned  with spherically symmetric velocity distributions
 and we will leave out the more realistic axially symmetric
 case \cite{HANSEN06a},\cite{HANSEN06b}. Such
 axially symmetric velocity distributions have, however, been found
 to have interesting consequences
 on the direct dark matter detection rates, especially in
 directional experiments \cite{Verg98}, \cite{Verg99}.\\
If the angular momentum dependence is ignored, by integrating Eq.
(\ref{distr.5}) one finds
                     \beq
\rho=4 \pi~\int f(E) \sqrt{2(E-\Phi(r))} dE
\label{distr.4}
\eeq
with the range of E as above. In this case one can obtain the density as a function of the potential.\\
Conversely if the density is given as a function of the potential
one can proceed to find the distribution function according to the
Eddington approach. The distribution then is a function of the
total energy$E=v^2/2+\Phi(r)$ and satisfies the Boltzmann Equation
with the collision term zero, i.e. \beq \left ({\bf
v}.\bigtriangledown _{{\bf r}}-\bigtriangledown \Phi.
\bigtriangledown _{{\bf v}} \right )f=0 \eeq In this case the
distribution can be expressed as follows:
\begin{equation}
f(E)=\frac{\sqrt{2}}{4 \pi^2}\frac{d}{dE} \int_E^0 \frac{d \Phi}{\sqrt{\Phi-E}}\frac{d \rho}{d \Phi}
\end{equation}
The above equation can be rewritten as:
\begin{equation}
f(E)=\frac{1}{2 \sqrt{2} \pi ^2} \left[ \int_E^0 \frac{d\Phi}{\sqrt{\Phi-E}} \frac{d^2 \rho}{d \Phi ^2}-
\frac{1}{\sqrt{-E}} \frac{d \rho}{d \Phi}|_{\Phi=0} \right ]
\end{equation}
Thus we can obtain the distribution, if the density $\rho$ is given as a function of the potential $\Phi$.
We find it convenient to rewrite  the last equation in terms of dimensionless variables by introducing:
\begin{equation}
\eta=\eta(\xi),~~ \eta=\frac{\rho}{\rho_0},~~ \xi=\frac{\Phi}{\Phi_0}
\label{dens1}
\end{equation}
The constants $\rho_0$ and $\Phi_0$, which set the scale of these two
 quantities, are related through Poisson's equation.
Then the Eddington distribution function takes the form:
\begin{equation}
f(e)=\frac{\rho_0}{\pi^2 |2 \Phi_0|^{3/2}}\left[\frac{\eta^{'}(0)}{\sqrt{e}}+
\int_0^e\frac{\eta^{''}(\xi)}{\sqrt{e-\xi}}d\xi \right]
\label{dens2}
\end{equation}
where $e$ is given by
\begin{equation}
e=-\frac{E}{|\Phi_0|}=\xi-\frac{\upsilon^2}{2 |\Phi_0|}
\label{dens3}
\end{equation}
i.e. $e$ the negative of the total energy $\epsilon$ ($e>0$. In the presence
of asymmetry one finds
\begin{equation}
e=-\epsilon=\xi-\frac{1}{2 |\Phi_0|}\left(\upsilon^2+\alpha_s \upsilon^2_t \right)
\label{dens3a}
\end{equation}
where $\alpha_s$ is the asymmetry parameter, $\upsilon_t$
is the tangential velocity. \\
For numerical integrations it is more convenient to rewrite the last integral as follows:
\begin{equation}
f(e)=\frac{\rho_0}{ \pi^2 |2 \Phi_0|^{3/2}}\left[\frac{\eta^{'}(0)}{\sqrt{e}}+ 2\eta^{''}(0)\sqrt{e}
+2\int_0^e \eta^{'''}(\xi)\sqrt{e-\xi}d\xi \right]
\label{dens4}
\end{equation}
The first term is singular as $e\rightarrow 0$. This, however, causes no problem,
 since, as we have seen in our earlier work, the integrals over the velocity distribution
are relevant in dark matter calculations and these remain finite as the velocity
 approaches the maximum velocity. Anyway in the example considered in the present work $\eta^{'}(0)=\eta^{''}(0)=0$.\\
 Once this function is known one obtains the velocity distribution of matter in our neighborhood
($r=r_s=0.8a~,a=$ the galactic radius)with respect to the center of the galaxy via the relation
 \beq
 f_{\upsilon}(v)=f(e)|_{r=r_s=0.8 a},
 \eeq
 which must be normalized. The characteristic feature of this approach is that the velocity distribution vanishes outside a given region specified by a cut off velocity $v_m$, by setting $e|_{r=r_s=0.8 a}=0$
\section{A Simple Realistic Density Profile}
We will consider three types of matter density:
\begin{itemize}
\item A  spherical ordinary matter density (bulge density)
\item Ordinary matter density in the form of a disc
\item Dark matter density.
\end{itemize}
\subsection{Spherical ordinary matter density}
\begin{figure}[!ht]
 \begin{center}
\rotatebox{90}{\hspace{-0.0cm} {$\frac{\rho(x)}{\rho_{0b}}\longrightarrow$}}
\includegraphics[scale=0.8]{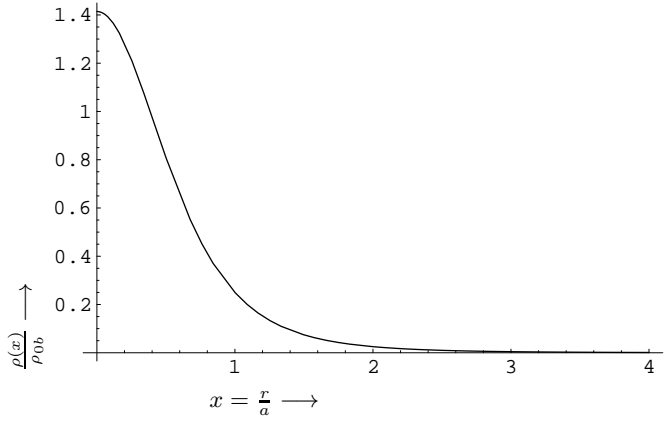}
{\hspace{-2.0cm}}\\
{\hspace{-2.0cm} {$x=\frac{r}{a}\longrightarrow$}}
 \caption{The ordinary matter density distribution in dimensionless units.}
 \label{gden}
  \end{center}
  \end{figure}

\begin{figure}[!ht]
 \begin{center}
\rotatebox{90}{\hspace{-0.0cm} {$\Phi_b(x)\longrightarrow 4 \pi G_N a^2 \rho_{0b}$}}
\includegraphics[scale=0.5]{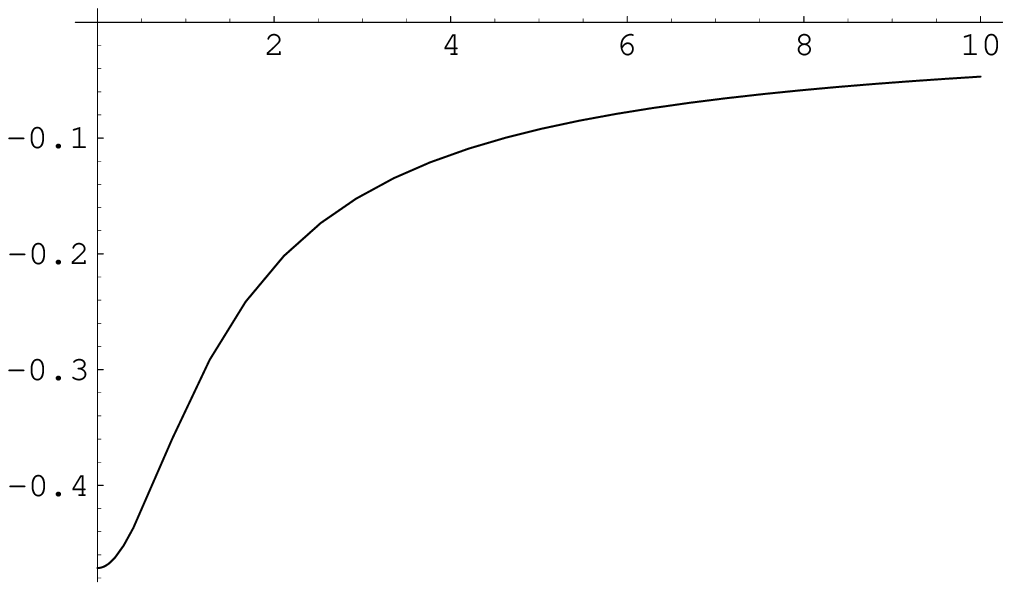}
\rotatebox{90}{\hspace{-0.0cm} {$v^2_b \longrightarrow4 \pi G_N a^2 \rho_{0b}$}}
\includegraphics[scale=0.5]{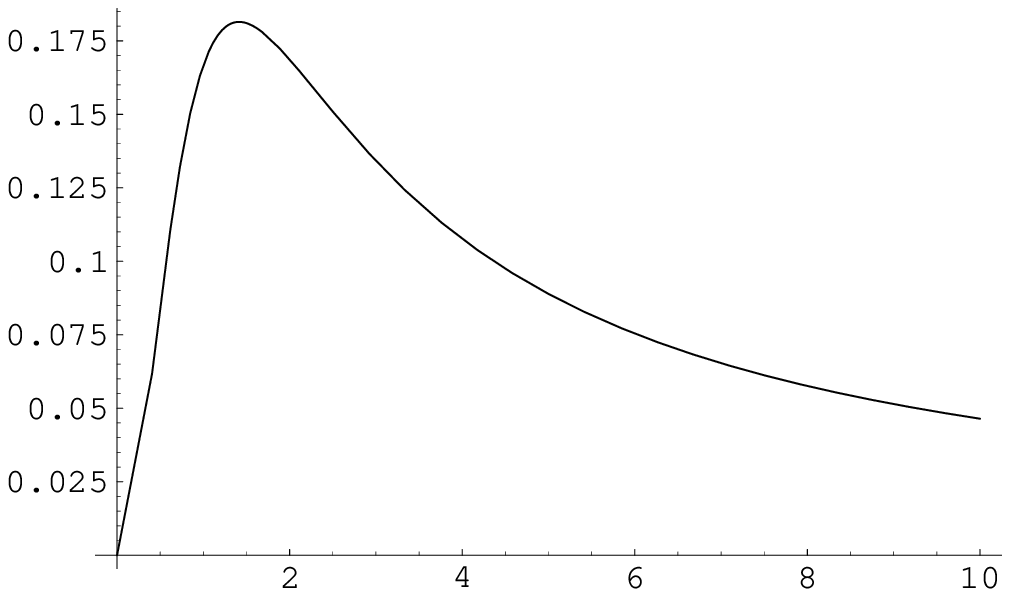}
{\hspace{-2.0cm}}\\
{\hspace{-2.0cm} {$x=\frac{r}{a}\longrightarrow$}}
 \caption{The Potential and the rotational velocity in units of $4 \pi G_N a^2 \rho_{0b}$ due to the spherical ordinary matter distribution discussed in the text.}
 \label{fig:bulge}
  \end{center}
  \end{figure}

To obtain analytical expressions we will simulate this density as
follows:
\begin{equation}
\rho_b (x)=\rho_{0b} \frac{\sqrt{2}}{(1+x^2)^{5/2}}~~,~~ x=\frac{r}{a}
\end{equation}
with $\rho_{0b}$ a constant and $a$ the galactic radius. So if
this were the whole story, $\rho_{0b}=2 \rho_s$ with $\rho_s$ the
mass density in our vicinity.  A distribution found by Plummer
\cite{PLUMMER} and Von Zeipel \cite{ZEIPEL},
$\rho(x)\propto(1+x^2)^{-5/2}$ was obtained by considering a gas
in spherical container with a specific heat ratio of $\gamma=1.2$
. We note that a value of $\gamma=1$, which leads to isothermal
distribution, was excluded  by observation. This normal matter
density profile is shown in Fig. \ref{gden}.
With this density one finds the potential:
\begin{equation}
\Phi_b (x)= -4 \pi G_N a^2 \rho_{0b}
\frac{\sqrt{2}}{6 \sqrt{x^2+1}}
\end{equation}
and the rotational velocity:
\begin{equation}
v^2_b (x)=x \frac{d \Phi(x)}{dx}= 4 \pi G_N a^2 \rho_{0b}
\frac{\sqrt{2} x^2}{3 \left(x^2+1\right)^{3/2}}
\end{equation}
The above potential and rotational velocity are shown in Fig. \ref{fig:bulge}.
\subsection{Ordinary matter density distributed on a disc}
\begin{figure}[!ht]
 \begin{center}
\rotatebox{90}{\hspace{-0.0cm} {$\Phi(x)\longrightarrow 4 \pi G_N a^2 \rho_{0d}$}}
\includegraphics[scale=0.8]{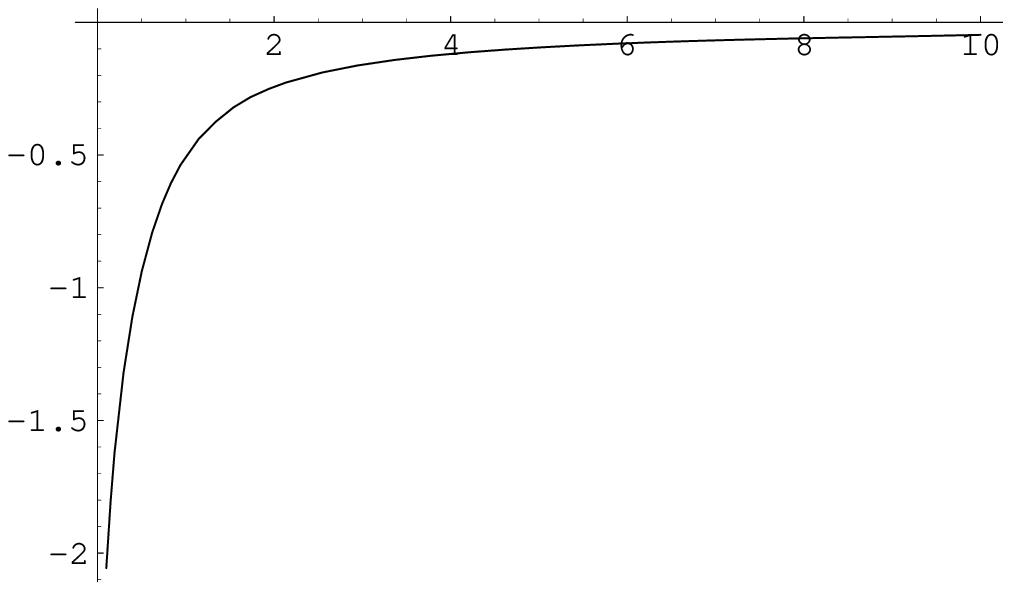}
{\hspace{-2.0cm}}\\
{\hspace{-2.0cm} {$x=\frac{r}{a}\longrightarrow$}}\\
\rotatebox{90}{\hspace{-0.0cm} {$v^2_d \longrightarrow4 \pi G_N a^2 \rho_{0d}$}}
\includegraphics[scale=0.65]{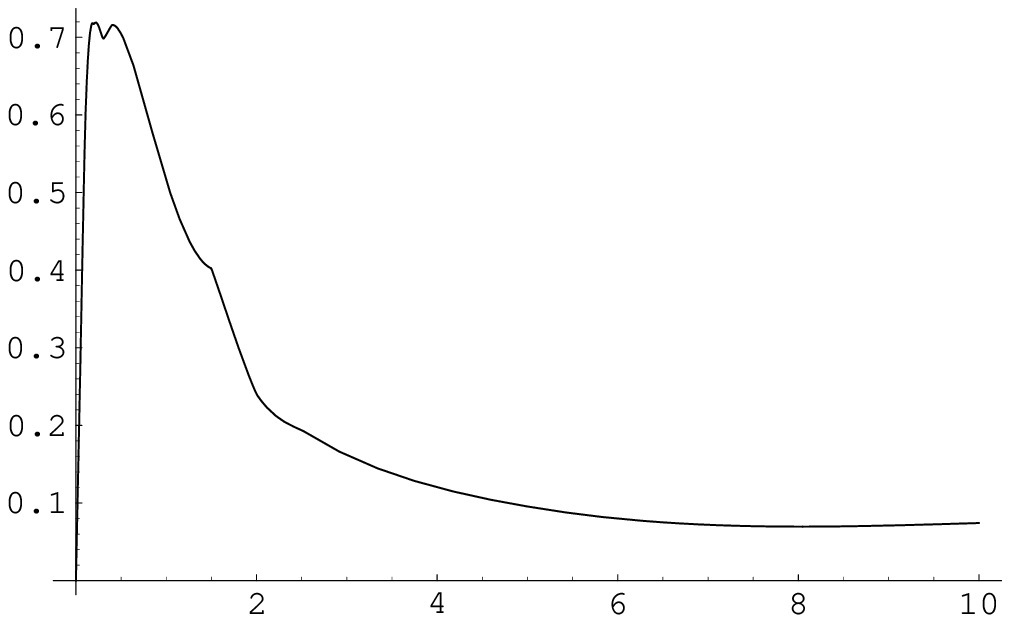}
\rotatebox{90}{\hspace{-0.0cm} {$v^2_{ed} \longrightarrow4 \pi G_N
a^2 \rho_{0ed}$}}
\includegraphics[scale=0.65]{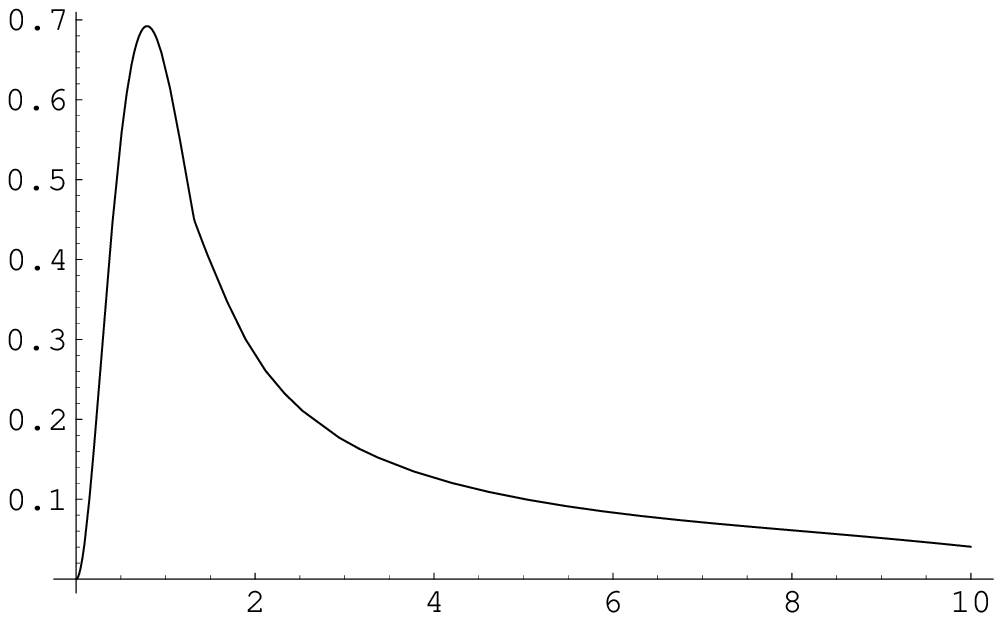}
{\hspace{-2.0cm}}\\
{\hspace{-2.0cm} {$x=\frac{r}{a}\longrightarrow$}}
 \caption{On top we show the Potential  in units of $4 \pi G_N a^2 \rho_{0d}$ on
 the plane of the galactic disc resulting from the ordinary matter distribution on the disc
 discussed in the text. At the bottom we show on the left the square of the rotational velocity
  resulting
 from  the above density and on the right the same quantity resulting from an exponential
 density profile.}
 \label{fig:disc}
  \end{center}
  \end{figure}

This is a more complicated problem. We will adopt the
simplification that the matter distribution is a $\delta$ function
along the axis of the disk. The radial density has been
successfully modelled  in the form of exponential profiles
\cite{LINPRING87}.  Nevertheless, to minimize the number of
parameters employed, we will assume further that it has the same
radial dependence as discussed above. In other words
\begin{equation}
\rho_d (x,z)=\rho_{0d} \delta(z)\frac{\sqrt{2}}{(1+x^2)^{5/2}}~~,~~ x=\frac{r}{a}
\end{equation}
where $r$ is now the radial distance from the axis of symmetry
(both $x$ and $z$ are measured in units of $a$). The potential now
takes the form:
\begin{equation}
\Phi_d(x,z)= 4 \pi G_N a^2 \rho_{0d} \frac{1}{\pi}
\int_0^{\infty } dk  \cos { (k z )} G(x, k)
\end{equation}
where the Green's function in momentum space is given in terms of the modified Bessel functions:
\begin{equation}
G (x,k)=G_{<} (x,k) +G_{>} (x,k)
\end{equation}
with
\begin{equation}
G_{>}=
-I_0(k x) \int_x^{\infty } K_0(k y) \rho_b (y) \, dy
\end{equation}
\begin{equation}
G_{<}=
-K_0(k x) \int_0^x I_0(k y) \rho_b (y) \, dy
\end{equation}
The radial rotational velocity  is given by
\begin{equation}
v_d^2(x,z)= 4 \pi G_N a^2 \rho_{0d} \frac{1}{\pi}
x \int_ 0^{\infty } \cos (k z) g(x,k) d k
   \end{equation}
with
\begin{equation}
g(x,k)=g_{<} (x,k) +g_{>} (x,k)
 \end{equation}
 \begin{equation}
g_{<}=
I_0(k x) K_0(k x) \rho_d (x)+k
   K_1(k x) \int_0^x I_0(k y) \rho_d (y) \, dy
\end{equation}
\begin{equation}
g_{<}=
I_0(k x) K_0(k x) \rho_d -k
   I_1(k x) \int_x^{\infty } K_0(ky) \rho_d (y) dy
\end{equation}
The obtained potential and the square of the rotational velocity
on the plane of the galactic plane are shown in
 Fig. \ref{fig:disc}. In the same Figure we also plot the same
 quantity obtained with the exponential profile:
 \begin{equation}
\rho_{ed} (x,z)=\rho_{0ed} \delta(z)e^{-3.5 x}~~,~~ x=\frac{r}{a}
\end{equation}
We see that the two densities give essentially the same rotational
velocities with the possible exception at very small distances.
\subsection{Dark matter density}
\begin{figure}[!ht]
 \begin{center}
\rotatebox{90}{\hspace{-0.0cm} {$\frac{\rho(x)}{\rho_0}\longrightarrow$}}
\includegraphics[scale=0.5]{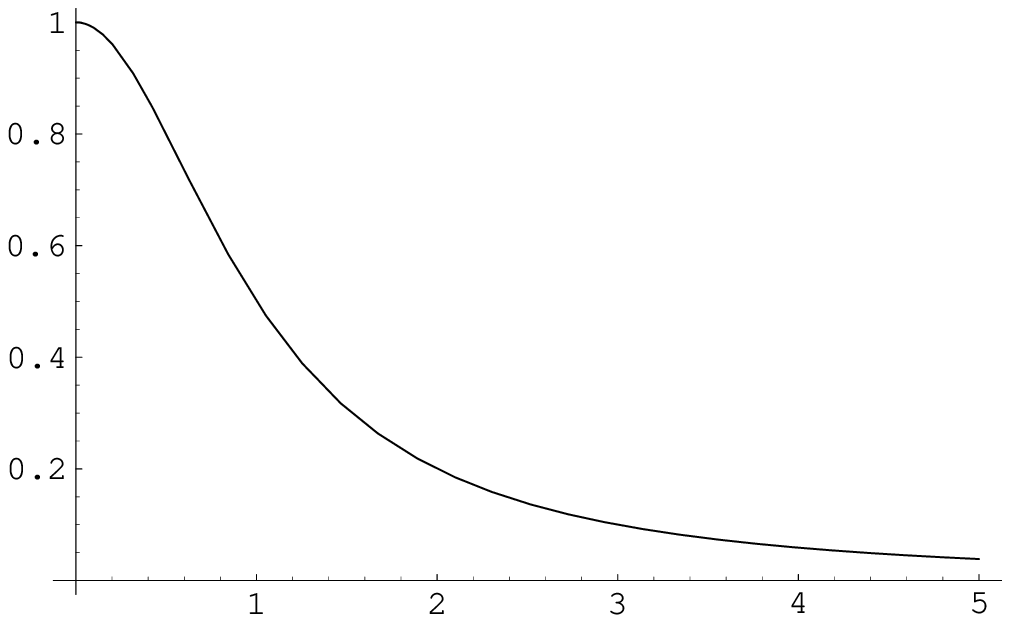}
\rotatebox{90}{\hspace{-0.0cm} {$\frac{\rho(x)}{\rho_0}\longrightarrow$}}
\includegraphics[scale=0.5]{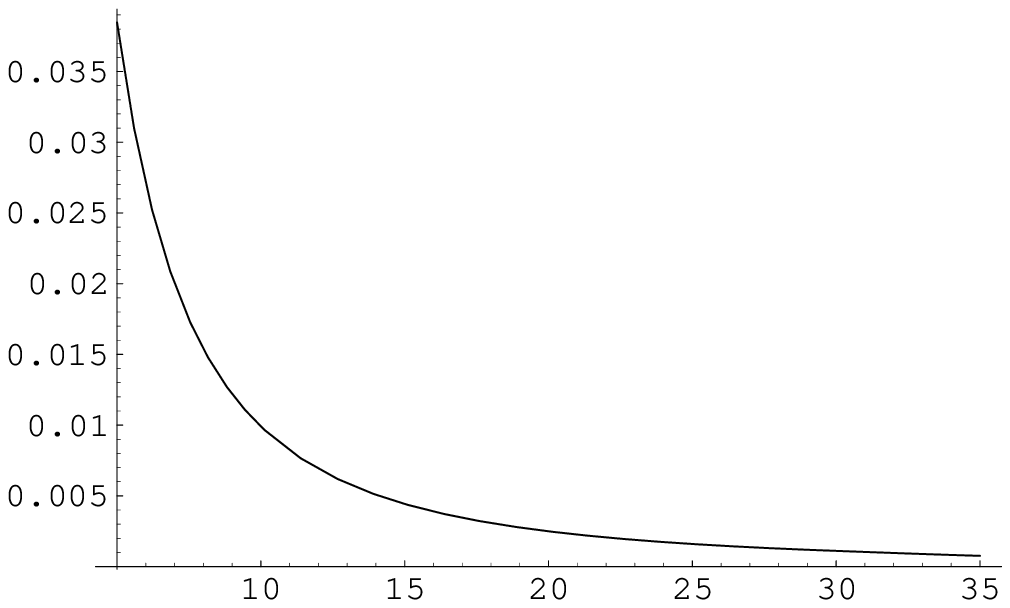}
{\hspace{-2.0cm}}\\
{\hspace{-2.0cm} {$x=\frac{r}{a}\longrightarrow$}}\\
\rotatebox{90}{\hspace{-0.0cm}
{$\frac{\rho(x)}{\rho_0}\longrightarrow$}}
\includegraphics[scale=0.8]{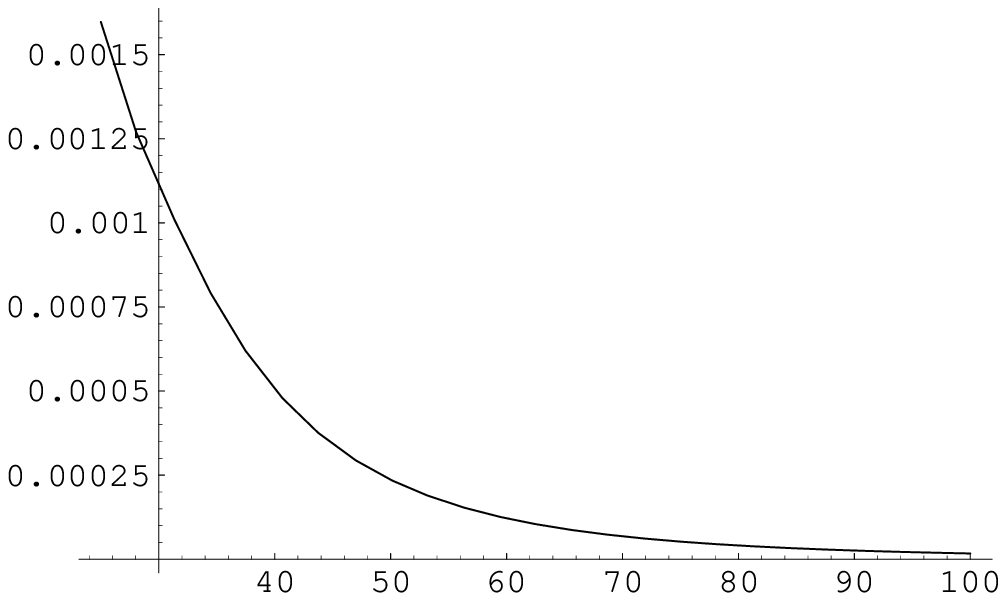}
{\hspace{-2.0cm}}\\
{\hspace{-2.0cm} {$x=\frac{r}{a}\longrightarrow$}}
 \caption{The dark matter density distribution in dimensionless units.}
 \label{density}
  \end{center}
  \end{figure}

\begin{figure}[!ht]
 \begin{center}
\rotatebox{90}{\hspace{-0.0cm} {$\xi=\frac{\Phi(x)}{\Phi_0}\longrightarrow $}}
\includegraphics[scale=0.5]{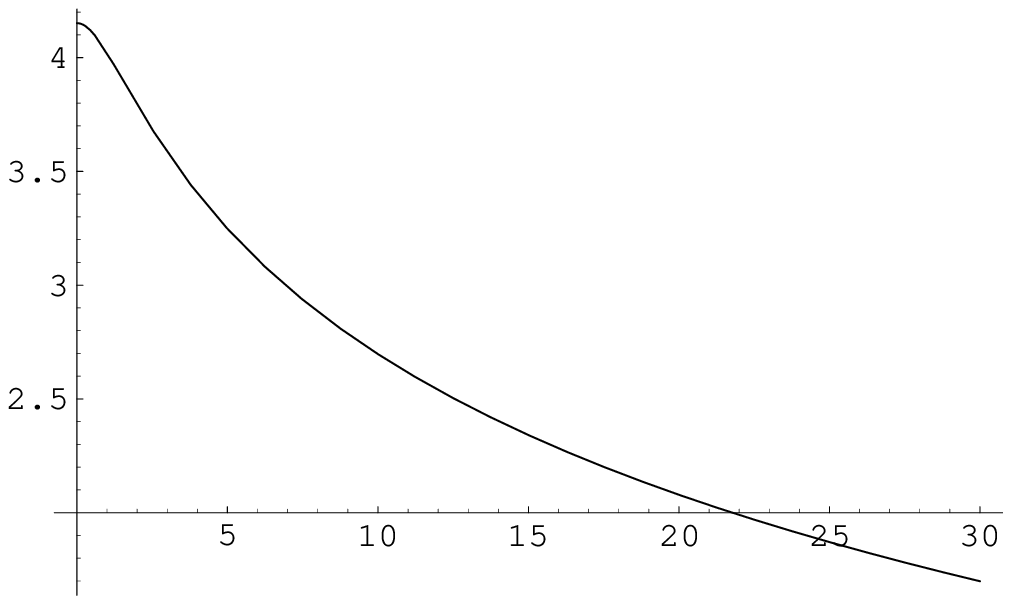}
\rotatebox{90}{\hspace{-0.0cm} {$\xi=\frac{\Phi(x)}{\Phi_0}\longrightarrow $}}
\includegraphics[scale=0.5]{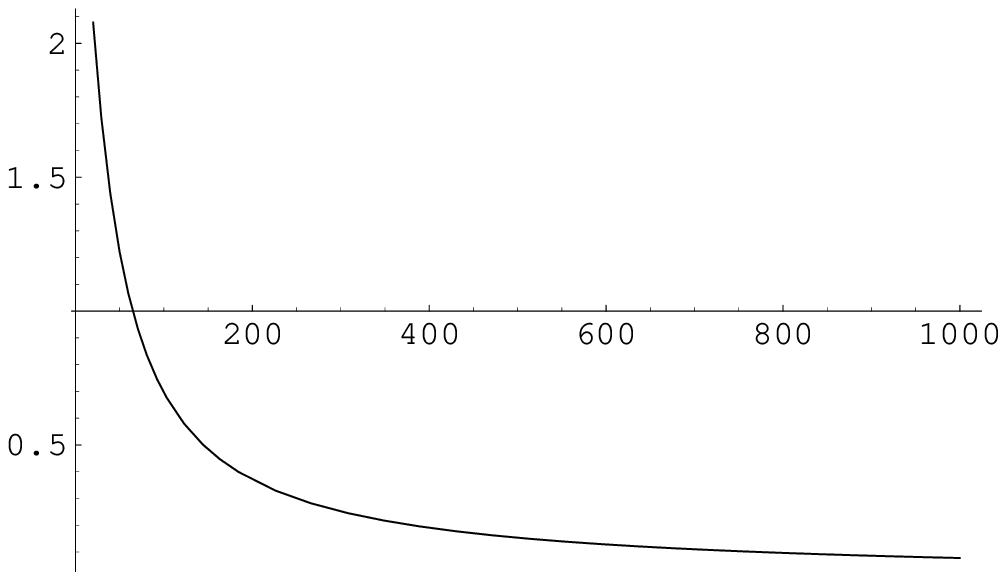}
{\hspace{-2.0cm}}\\
{\hspace{0.2cm} {$x=\frac{r}{a}\longrightarrow$}}
 \caption{The potential $\xi=\frac{\Phi(x)}{\Phi_0}$, 'i.e the absolute value of the potential
 in units $4 \pi G_Na^2 \rho_0 $, obtained with the density function shown in Fig. \ref{density}
 and chosen to vanish at infinity. The potential drops to zero very slowly at large distances.}
 \label{potential}
  \end{center}
  \end{figure}

\begin{figure}[!ht]
 \begin{center}
\rotatebox{90}{\hspace{-0.0cm} {$\eta \longrightarrow$}}
\includegraphics[scale=0.8]{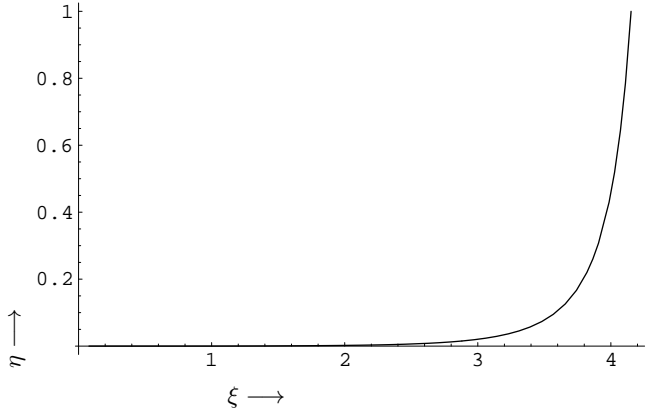}
{\hspace{-2.0cm}}\\
{\hspace{-2.0cm} $\xi \longrightarrow$}
 \caption{The density $\eta$ as a function of the potential $\xi$}
 \label{denpot}
  \end{center}
  \end{figure}

\begin{figure}[!ht]
 \begin{center}
\rotatebox{90}{\hspace{-0.0cm} {$v_{rot}\longrightarrow \sqrt{4\pi G_Na^2 \rho_0}$}}
\includegraphics[scale=0.5]{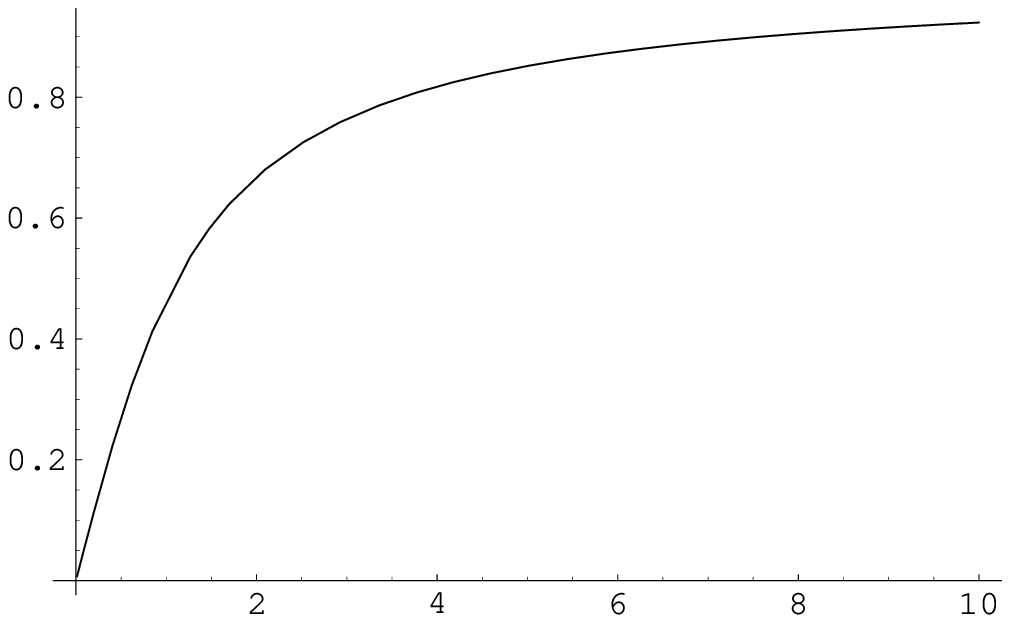}
 { \rotatebox{90}{\hspace{-0.0cm} {$v_{rot}\longrightarrow
\sqrt{4\pi G_Na^2 \rho_0}$}}}
\includegraphics[scale=0.5]{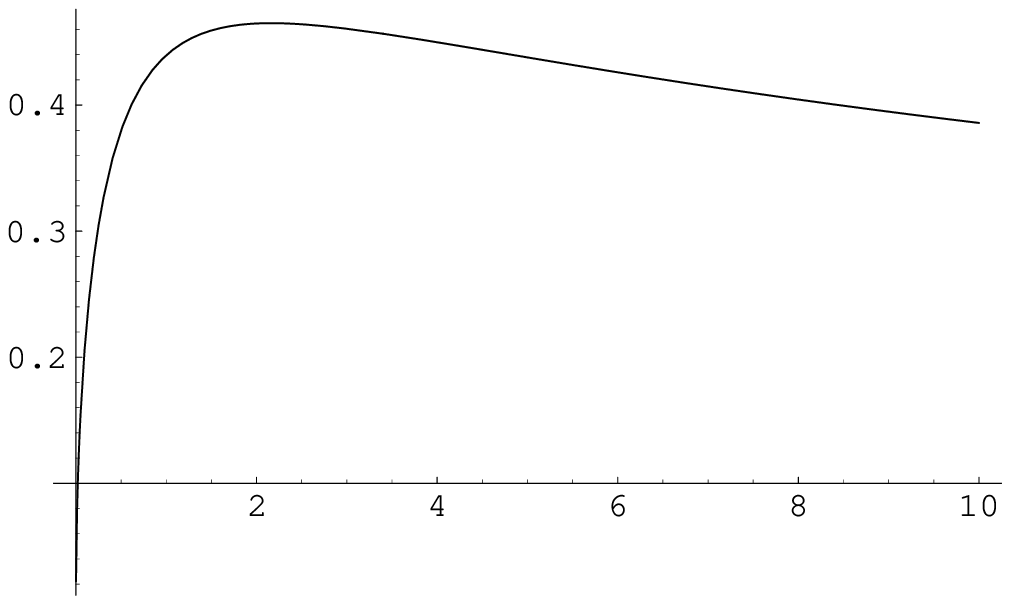}
{\hspace{-2.0cm}}\\
{\hspace{-0.0cm} $x \longrightarrow \frac{x}{a}$}
 \caption{The rotational velocity due to dark matter as a function of the distance in units
 of $\sqrt{4\pi G_Na^2 \rho_0}$. Shown on the left is the one obtained with the
  density profile of Eq. (\ref{dens5}) adopted in this work, while on the right the
  NFW profile \cite{ULLIO},
  see Eq. (\ref{dens5b}), was employed.}
 \label{rotvel}
  \end{center}
  \end{figure}

There are many halo density profiles, which have been employed. In
the present case we will consider only spherical distributions,
since it is not easy to extend the Eddington approach to deal with
the most general case. Among the most commonly used are:
\begin{itemize}
\item The simple density profile
\begin{equation}
\rho(x)=\frac{\rho_0}{1+x^2},~~ x=\frac{r}{a}
\label{dens5}
\end{equation}
with $a$ the radius of the Galaxy. This profile has the advantage
that the rotational velocity remains constant with the distance
from the center of the galaxy becomes very large.
 \item Another simple profile is:
\begin{equation}
\rho(x)=\frac{\rho_0}{x(1+x)^2},~~ x=\frac{r}{a}
 \label{dens5b}
\end{equation}
suggested by N-body simulations \cite{ULLIO}. This profile
provides a better description of the expected density near the
center of the galaxy. It does not, however, predict the constancy
of the rotational velocities at large distances.
 \end{itemize}
 In the present work we will consider the density profile of Eq.
 (\ref{dens5})\\
Unfortunately with this density the potential diverges at
infinity. On the other hand the solution to Poisson's equation is
finite at the origin and it can be chosen to vanish there, i.e. it
takes the form:
\begin{equation}
\frac{\Phi}{\Phi_0}=\frac{tan^{-1}(x)}{x}+\frac{1}{2}\ln(1+x^2)-1
\label{dens6}
\end{equation}
One may choose a radius $x=c$ outside of which the density can be chosen to go faster to zero. One convenient choice is:
\beq
\rho(x)=c_2 \frac{\rho_0}{(1+x^2)^2}+c_3 \frac{\rho_0}{(1+x^2)^3}
\eeq
with the requirement that at $x=c$ the density is continuous with a continuous derivative.
we thus find:
 \beq
 \rho_{>}(x)=\rho_0\left[
\frac{2
   \left(c^2+1\right)}{\left(x^2+1\right)^2}-\frac{\left(c^2+1\right)^2}{\left(x^2+1\right)^3} \right]
   \eeq
   The obtained density is shown in Fig. \ref{density}.
   The potential in the region $x\prec c$ is the same as before:
\beq
\frac{\Phi_{<}(x)}{\Phi_0}=I_{<}(x)+c_3
   \eeq
The solution in the outer region takes the form:
   \beq
   \frac{\Phi_{>}(x)}{\Phi_0}=I_{<}(c)+I_{>}(x)-I_{>}(c)-c_3
   \eeq
   where the constant $c_3$ can be chosen to make the potential vanish at some point and
   \beq
I_{<}=
\frac{\tan ^{-1}(x)}{x}+\frac{1}{2} \log
   \left(x^2+1\right)
   \eeq
\beq
I_{>}=
\frac{1}{8}
   \left(\frac{\left(c^2+1\right)^2}{x^2+1}+\frac{\left
   (c^2-7\right) \tan ^{-1}(x)
   \left(c^2+1\right)}{x}+\frac{\left(-c^4+6
   c^2+15\right) \tan ^{-1}(c)-c
   \left(c^2+15\right)}{x}\right)
  \eeq
  by choosing $c_3$ as
  $$c_3= \frac{1}{4} \left(2 \log \left(c^2+1\right)+7\right)$$
the potential can be made to vanish at infinity. It is shown in Fig. \ref{potential}.
  With the above ingredients it is not very difficult to obtain the function needed in the Eddington approach,
  namely $\eta(\xi)$. The results are shown in Fig. \ref{denpot}.
In all cases Poisson's equation yields a relation  between $\Phi_0$ and $\rho_0$, namely
$$\Phi_0=-4\pi G_Na^2 \rho_0.$$
The obtained rotational velocity curve, due to dark mater alone, is given in Fig. \ref{rotvel} in units of
$\sqrt{4\pi G_Na^2 \rho_0}$
From the observed rotational velocity one can fix the constant $\rho_0$.
\subsection{Combining ordinary matter and dark matter}
\begin{figure}[!ht]
 \begin{center}
\rotatebox{90}{\hspace{-0.0cm} {$\Phi(x)\longrightarrow 4 \pi G_N a^2 ~\rho_{0}$}}
\includegraphics[scale=0.5]{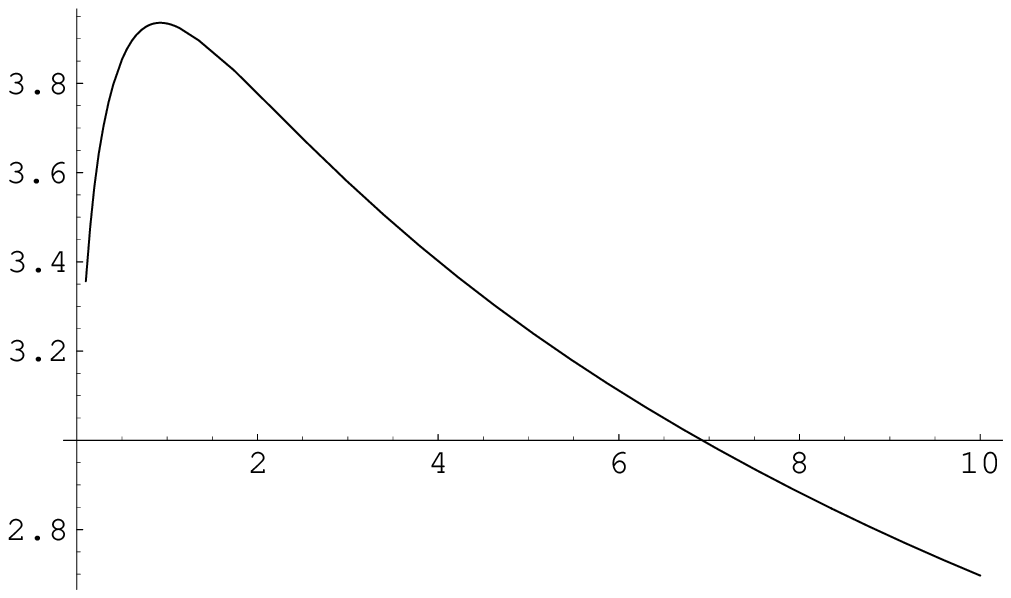}
  \rotatebox{90}{\hspace{-0.0cm} {$\Phi(x)\longrightarrow 4 \pi
G_N a^2 ~\rho_{0}$}}
\includegraphics[scale=0.5]{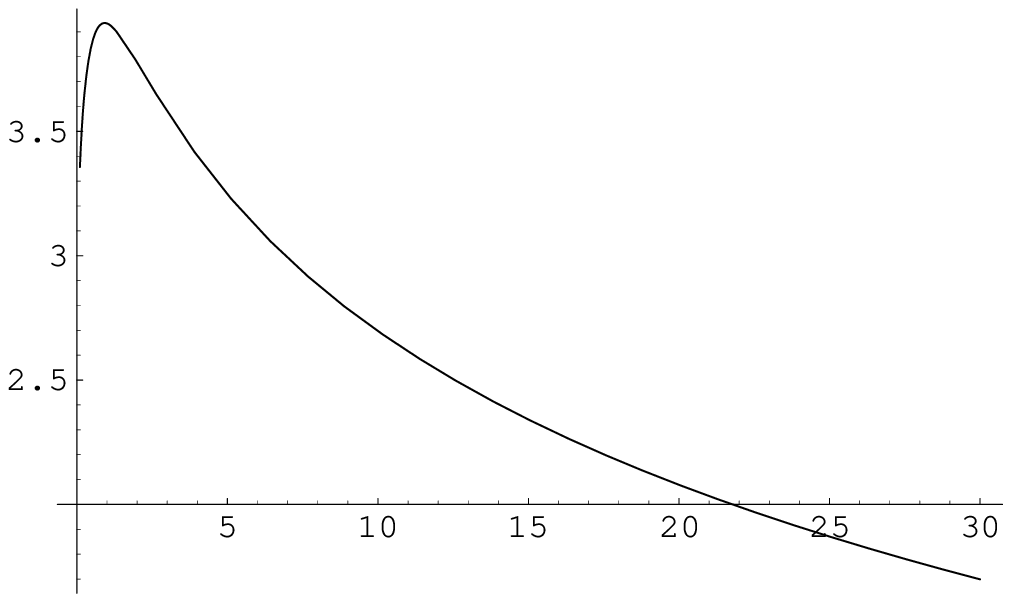}
{\hspace{-2.0cm}}\\
{\hspace{-0.0cm} {$x=\frac{r}{a}\longrightarrow$}}
 \caption{The Potential due to all three matter distributions considered in the text.}
 \label{fig:cdmdiscbulge1}
  \end{center}
  \end{figure}

We will now combine the three kinds of distribution considered above. We will assume that $\rho_{0d}=\rho_{ob}=\rho_s$ and $\rho_s=\rho_0$. This means that in our vicinity the density of dark matter is equal to that of ordinary matter. Furthermore the ordinary matter density in our vicinity is equally split between the spherical and disc geometries. The obtained results are shown in Fig. \ref{fig:cdmdiscbulge1}-\ref{fig:cdmdiscbulge1}.
        \begin{figure}[!ht]
 \begin{center}
\rotatebox{90}{\hspace{-0.0cm} {$\upsilon_{rot}\longrightarrow \sqrt{4 \pi G_N a^2 ~\rho_{0}}$}}
\includegraphics[scale=0.5]{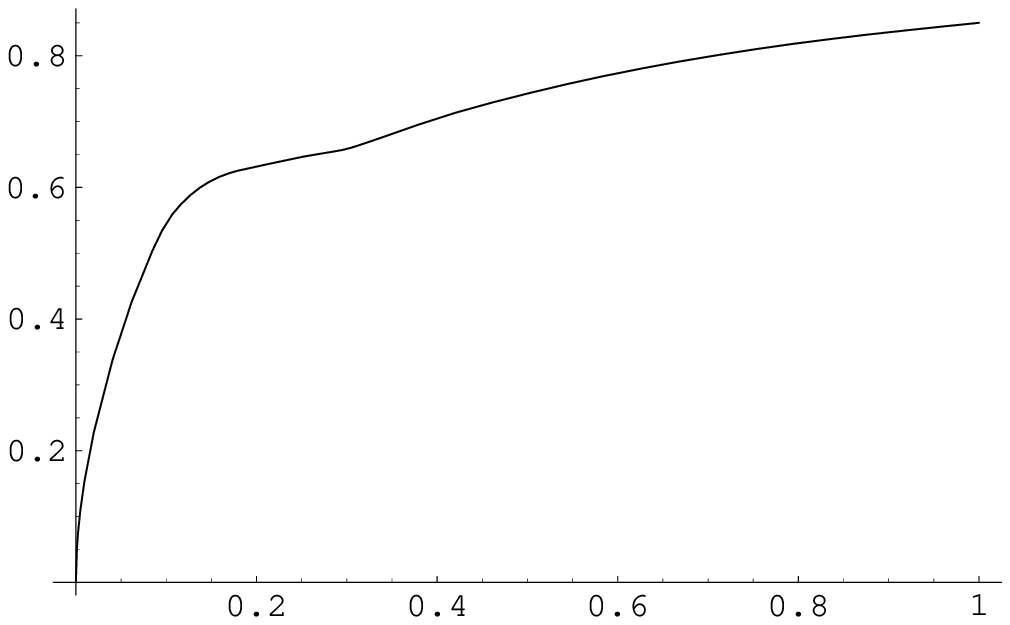}
 { \rotatebox{90}{\hspace{-0.0cm} {$\upsilon_{rot}\longrightarrow
\sqrt{4 \pi G_N a^2 ~\rho_{0}}$}}}
\includegraphics[scale=0.5]{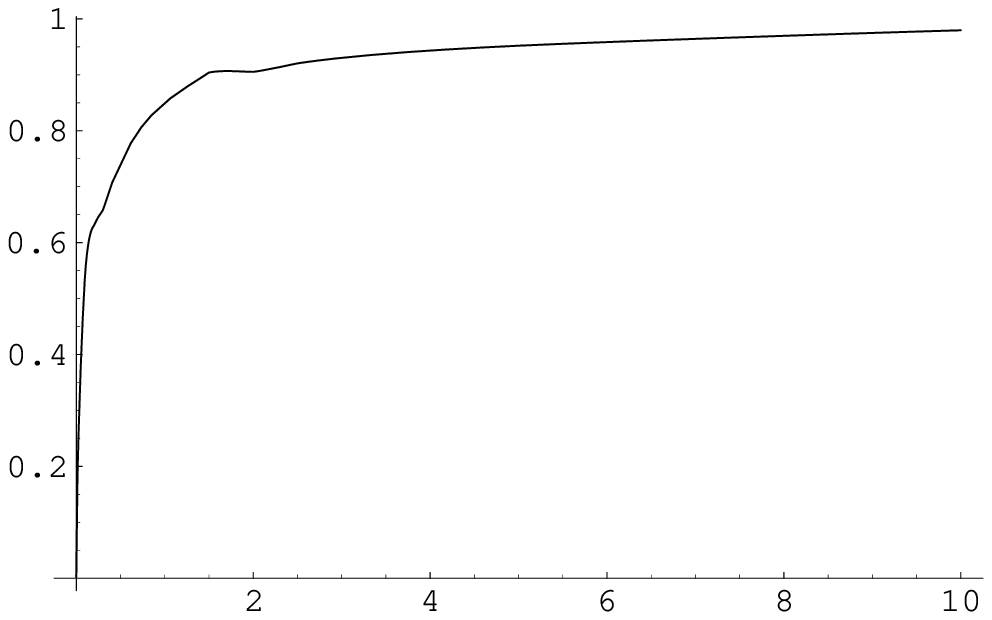}
{\hspace{-2.0cm}}\\
{\hspace{-0.0cm} {$x=\frac{r}{a}\longrightarrow$}}
 \caption{The rotational velocity resulting from the potential of Fig. \ref{fig:cdmdiscbulge1}.}
 \label{fig:cdmdiscbulge2}
  \end{center}
  \end{figure}
  We see that in order to fit the rotational velocity of the sun, $220~km/s$ we need a density $\rho_0=1.0\times 10^{-21}kg/m^3$. This is in agreement with the value of $0.3~GeV/cm^3=0.5\times 10^{-21}~kg/m^3$ used in calculations of the rates for direct dark matter searches (in our model the dark component is half of the total).
\section{The velocity distribution in our Vicinity}
It is clear that the velocity distribution and, in particular the maximum allowed velocity, is related to the escape velocity via the density $\rho_0$. For the moment we will ignore the asymmetric term and set $\alpha_s=0$.
\subsection{Dark Matter only}
\begin{figure}[!ht]
 \begin{center}
\rotatebox{90}{\hspace{-0.0cm} {$f_{\upsilon}(y)\longrightarrow$}}
\includegraphics[scale=0.5]{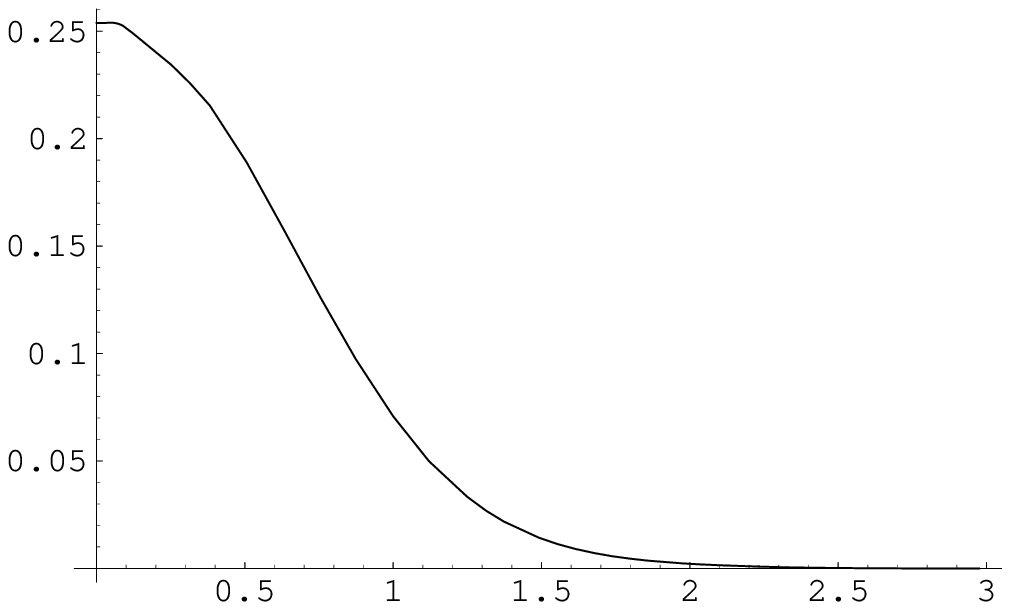}
  \rotatebox{90}{\hspace{-0.0cm} {$y^2
f_{\upsilon}(y)\longrightarrow$}}
\includegraphics[scale=0.5]{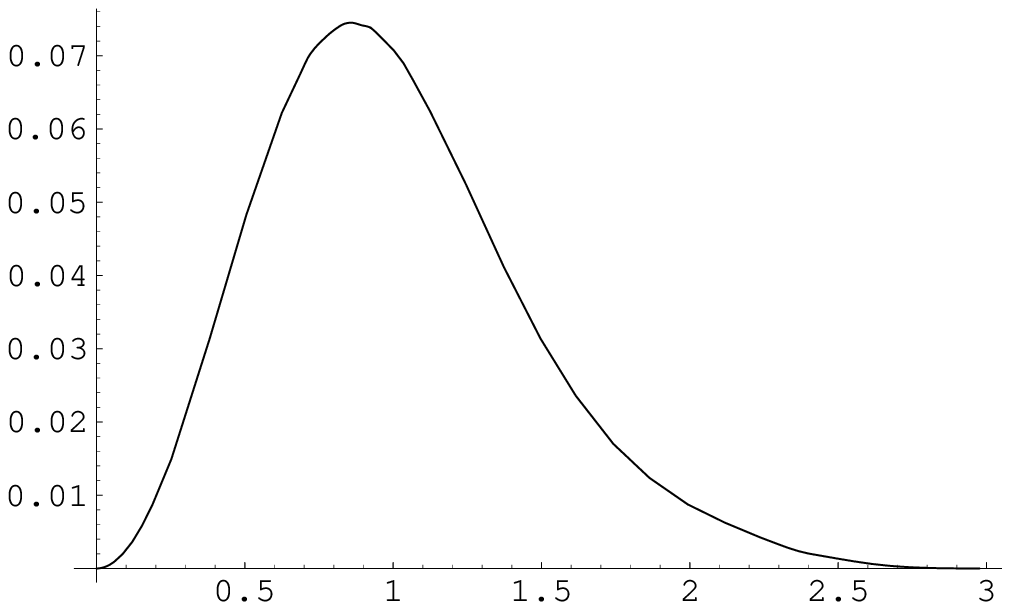}
{\hspace{-2.0cm}}\\
{\hspace{-2.0cm}}\\
{\hspace{-0.0cm} $y\longrightarrow$}
 \caption{The velocity distributions $f_{\upsilon}(y)$ and $y^2 f_{\upsilon}(y)$ in units of
  the parameter $\sqrt{4\pi G_Na^2 \rho_0}$, i.e. $y=v/\sqrt{4\pi G_Na^2 \rho_0}$
   obtained with
 the density profile of Eq. (\ref{dens5})}
 \label{veldis}
  \end{center}
  \end{figure}

\begin{figure}[!ht]
 \begin{center}
\rotatebox{90}{\hspace{-0.0cm} {$f_{\upsilon}(y)\longrightarrow$}}
\includegraphics[scale=0.5]{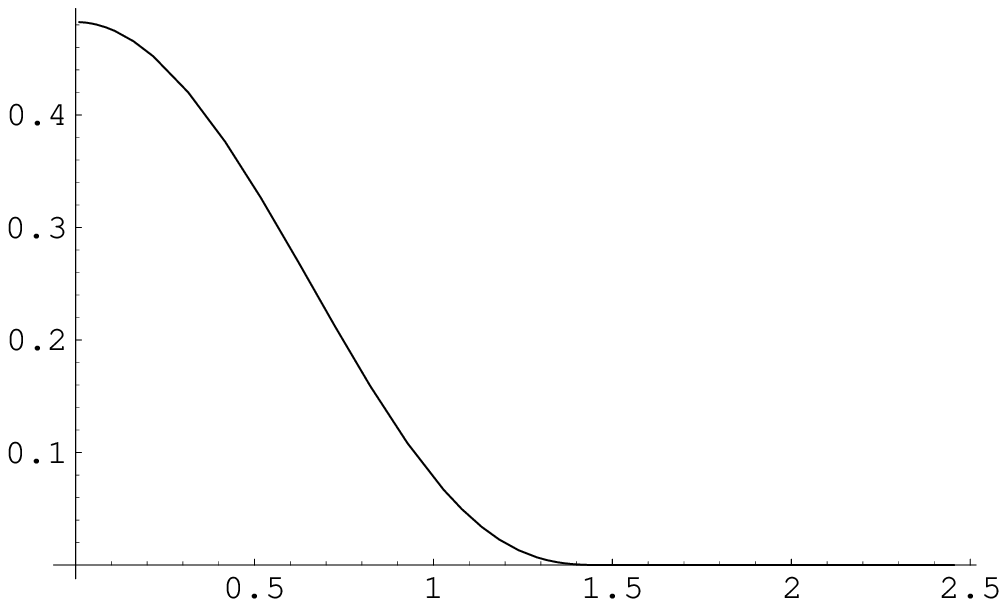}
  \rotatebox{90}{\hspace{-0.0cm} {$y^2
f_{\upsilon}(y)\longrightarrow$}}
\includegraphics[scale=0.5]{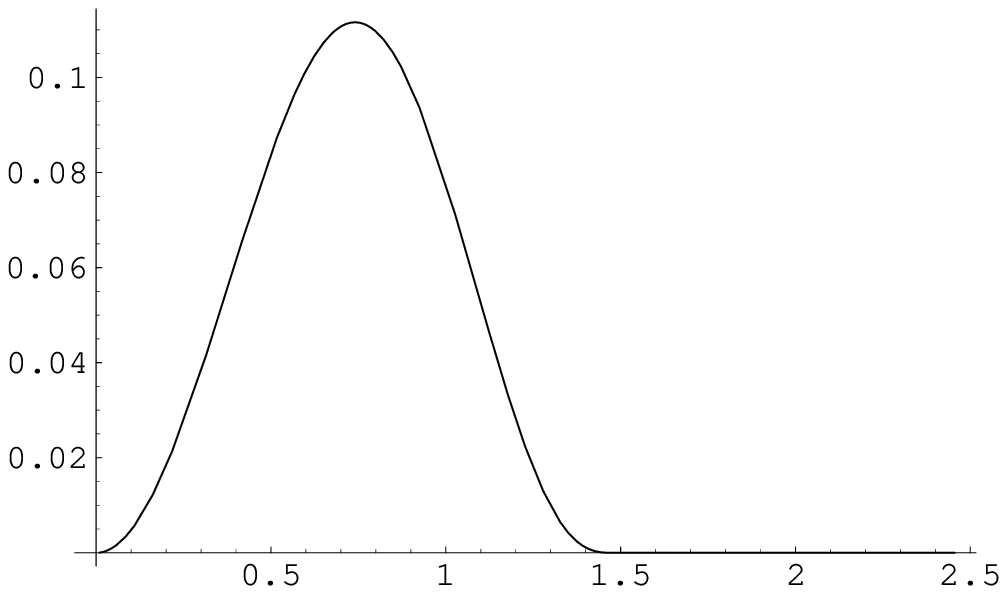}
{\hspace{-2.0cm}}\\
{\hspace{-2.0cm}}\\
{\hspace{-0.0cm} $y\longrightarrow$}
 \caption{The velocity distributions $f_{\upsilon}(y)$ and $y^2 f_{\upsilon}(y)$ in units of
 the parameter $\sqrt{4\pi G_Na^2 \rho_0}$, i.e. $y=v/\sqrt{4\pi G_Na^2 \rho_0}$, obtained with
 the density profile of Eq. (\ref{dens5b})}
 \label{NFW}
  \end{center}
  \end{figure}

\begin{figure}[!ht]
 \begin{center}
\rotatebox{90}{\hspace{-0.0cm} {$f_{\upsilon}(y)\longrightarrow$}}
\includegraphics[scale=0.5]{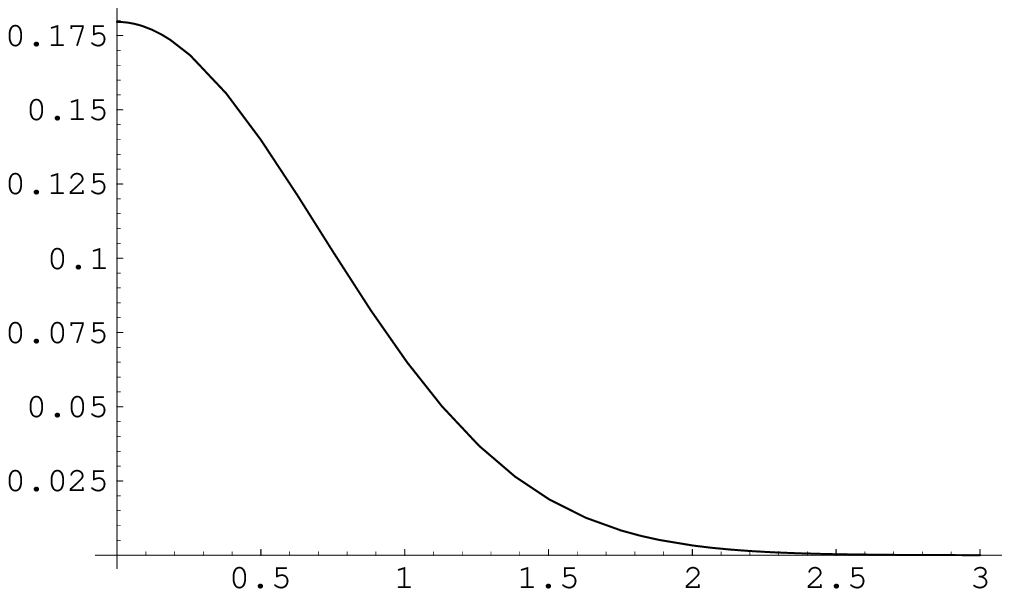}
  \rotatebox{90}{\hspace{-0.0cm} {$y^2
f_{\upsilon}(y)\longrightarrow$}}
\includegraphics[scale=0.5]{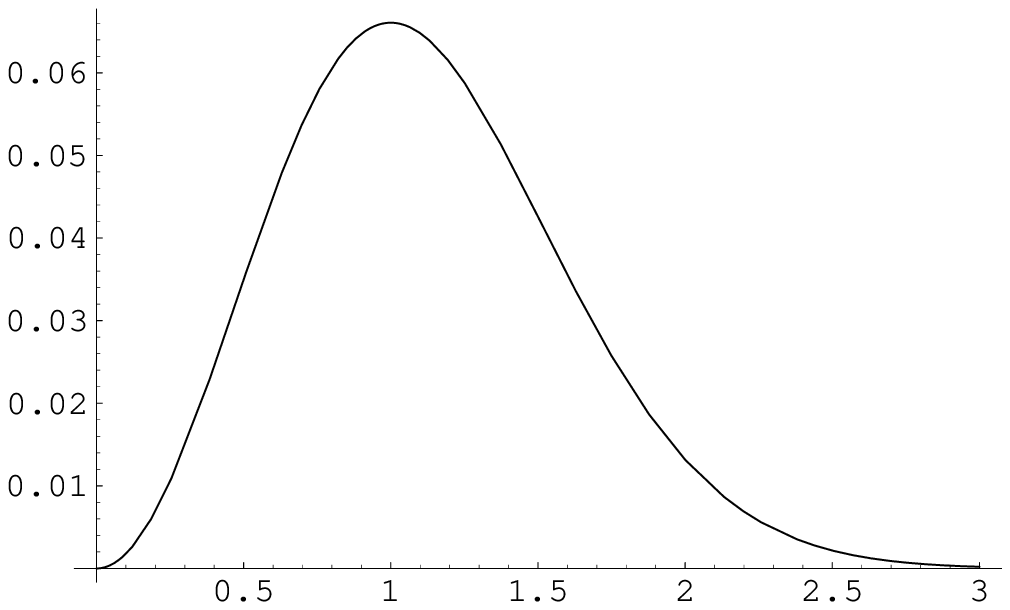}
{\hspace{-2.0cm}}\\
{\hspace{-2.0cm}}\\
{\hspace{-0.0cm} $y\longrightarrow$}
 \caption{The same as in Fig. \ref{veldis} in the case of the M-B distribution. Now, however,
 the parameter
$y$ is the velocity in units of the sun's rotational velocity, i.e. $y=v/v_0$ with
$v_0=2.2 \times 10^5 m/s$. Note that, in these units, the escape velocity is at $2.84$ }
 \label{MBvel}
  \end{center}
  \end{figure}

The relation between the density and the potential has already
been shown in Fig. \ref{denpot}. In this case we obtain the
velocity distribution shown in Fig. \ref{veldis}, with the usual
normalization imposed $$\int_0^{y_m} y^2 f_{\upsilon}(y)dy=1.$$
Instead of the velocity we have used the dimensionless quantity y:
$$y=\frac{v}{\sqrt{4 \pi G_Na^2 \rho_0}}.$$
  For comparison we present the velocity distributions obtained
  with the profile of Eq. (\ref{dens5b}) in Fig. \ref{NFW} and with
  the standard M-B distribution in Fig. \ref{MBvel}.
The maximum velocity allowed by our distribution is only $y_m=2.8$. Assuming $a=3.1 \times 10 ^{20}~m$ and $\rho|_{x=1}=\rho_s/2$, i.e. half of the local density to be due to dark matter, we find
$\rho_0=\rho_s$
 which yields $v_m=y_m \sqrt{4 \pi G_N a^2 \rho_0}=2.8\times 270 km/s=7.5\times 10^5 m/s$. This is  a bit higher than the escape velocity,  $v_{esc}=6.2 \times 10^5 m/s$, assumed in theories employing the M-B distribution.
For comparison we present the same quantities for the M-B distribution in Figs \ref{MBvel}-\ref{MBvel}.
Since the M-B distribution does not go to zero at finite values of the velocity,
 the maximum allowed velocity  is set by hand equal to the escape velocity.\\
 We should emphasize that in this approach we encounter two characteristic velocities. One is the rotational velocity and the other the maximum allowed velocity $v_m$. The first depends on the square root of the derivative of the potential, while the second scales with the square root of the potential itself.
\subsection{Both ordinary matter and dark matter}
\begin{figure}
 \begin{center}
\rotatebox{90}{\hspace{-0.0cm} {$\eta \longrightarrow$}}
\includegraphics[scale=0.8]{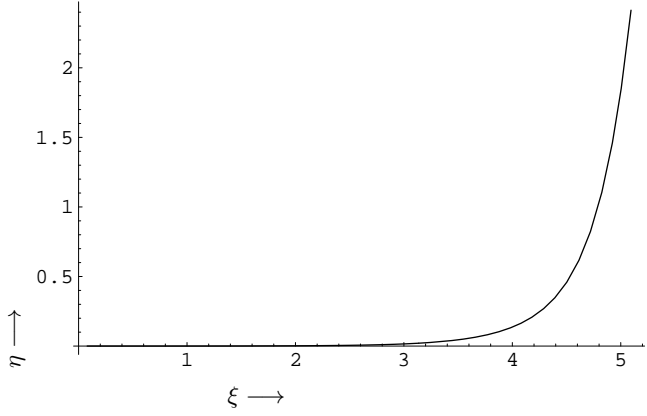}
{\hspace{-2.0cm}}\\
{\hspace{-2.0cm} $\xi \longrightarrow$}
\caption{The density $\eta$ as a function of the potential $\xi$, when both ordinary and dark matter are included.}
 \label{gdenpot}
  \end{center}
  \end{figure}

To improve the situation one will attempt to include gravity.
Since it is very hard to incorporate the disc geometry into the
Eddington approach, we will attempt to mimic the gravitational
effects of ordinary matter with a spherical distribution like the
one discussed above, but twice a large, so that in our position
the usual and cold dark matter contribution are about equal. The
thus obtained function $\eta=\eta(\xi)$ is shown in Fig.
\ref{gdenpot}.

\begin{figure}[!ht]
 \begin{center}
\rotatebox{90}{\hspace{-0.0cm} {$f_{\upsilon}(y)\longrightarrow$}}
\includegraphics[scale=0.5]{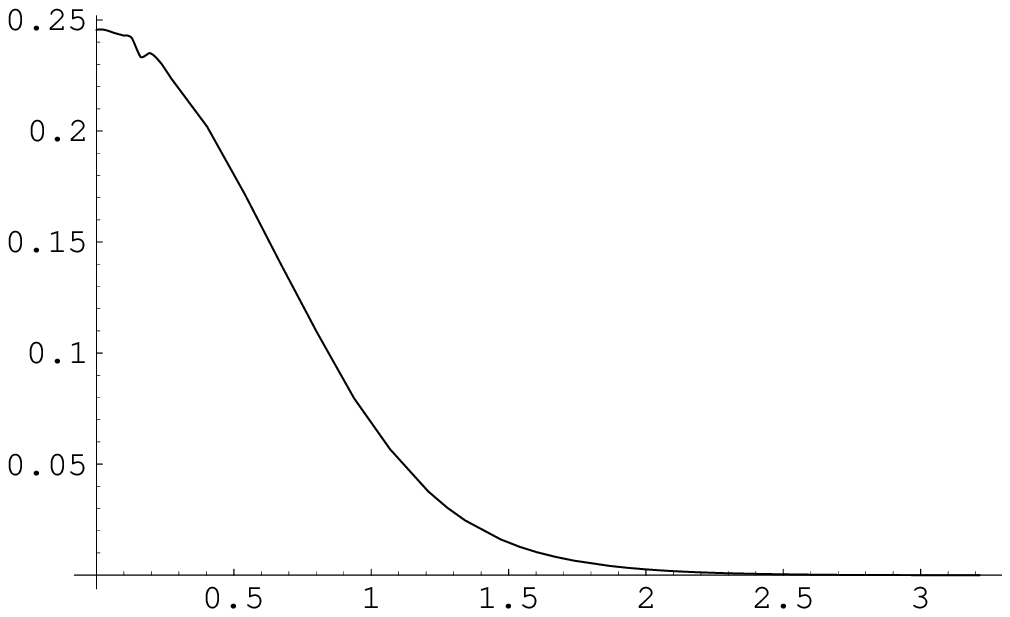}
  \rotatebox{90}{\hspace{-0.0cm} {$y^2
f_{\upsilon}(y)\longrightarrow$}}
\includegraphics[scale=0.5]{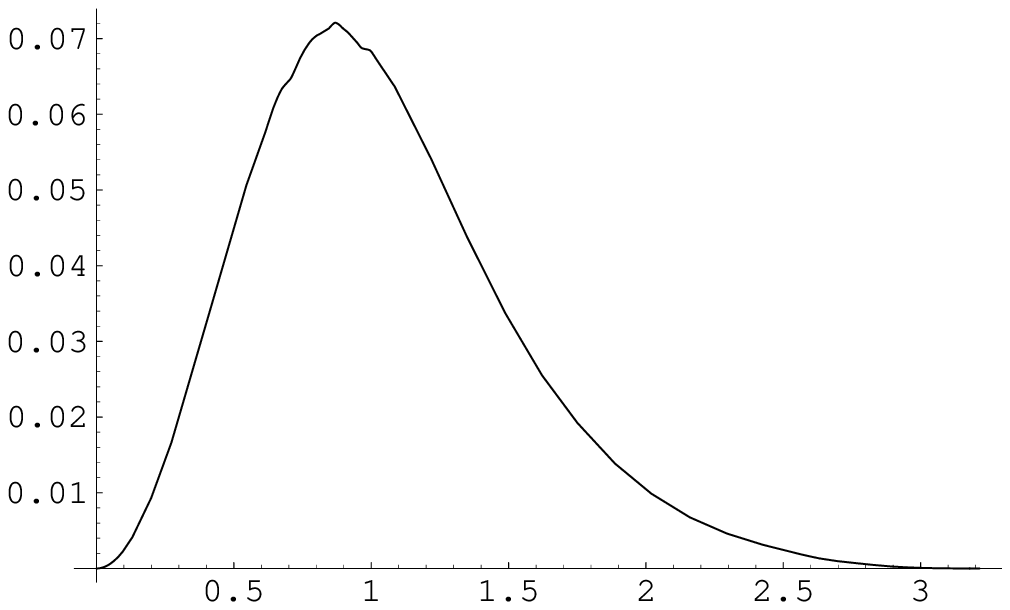}
{\hspace{-2.0cm}}\\
{\hspace{-2.0cm}}\\
{\hspace{-0.0cm} $y\longrightarrow$}
 \caption{The same as in Fig. \ref{veldis}, when both ordinary and dark matter are included.}
 \label{gveldisboth}
  \end{center}
  \end{figure}

 In this case we find $y_m=3.2$. In other words the effect of ordinary matter is small, since the dark matter
 potential in our vicinity is about  $12$ times stronger than the potential due to ordinary matter.
 Again the condition $\rho|_{x=1}=\rho_s$ implies that overall constant $\rho_0$ in the density distribution
 is $\rho_s$, which gives $v_m=3.1\times  2.7\times 10^5~m/s=8.0\times 10^5~m/s$.
With this modification the obtained  results for the velocity distribution are shown in Figs \ref{gveldisboth}.  We see that the inclusion of gravity
has very little effect on the velocity distribution. So  in what follows we will consider only
 the dark matter component.
  \section{Direct dark matter rates}
  In the present work  we find it convenient to write the event rate in the form \footnote { see our previous
recent work \cite{JDV06,JDV05} for a more detailed discussion and additional references} :
\begin{equation}
R= \bar{K} \left[c_{coh}(A,\mu_r(A)) \sigma_{p,\chi^0}^{S}+
c_{spin}(A,\mu_r(A))\sigma _{p,\chi^0}^{spin}~\zeta_{spin} \right]
\label{snew}
\end{equation}
In this expression $\sigma^S_{N,\chi^0}$ is the LSP-nucleon scalar
cross section, while $\sigma^{spin}_{p,\chi^0}$ is the proton
cross section associated with the spin. The quantity
$\zeta_{spin}$is given by \cite{JDV06}:
\begin{equation}
\zeta_{spin}= \frac{1}{3(1+\frac{f^0_A}{f^1_A})^2}S(u)
\label{2.10a}
\end{equation}
\begin{equation}
S(u)\approx S(0)=[(\frac{f^0_A}{f^1_A} \Omega_0(0))^2
  +  2\frac{f^0_A}{ f^1_A} \Omega_0(0) \Omega_1(0)+  \Omega_1(0))^2  \, ]
\label{s(u)}
 \end{equation}
 The static spin matrix elements are obtained in the context of a given nuclear
 model (see, e.g., previous work \cite{Ress}$-$\cite{DIVA00},\cite{JDV06,JDV05} and references therein). Even, though most of what we are going to say applies in
 the case of the spin induced rate we are not going to further
 elaborate here.

\begin{figure}[!ht]
 \begin{center}
\rotatebox{90}{\hspace{-0.0cm} {$\Psi(a \sqrt{u}) \longrightarrow$}}
\includegraphics[scale=0.6]{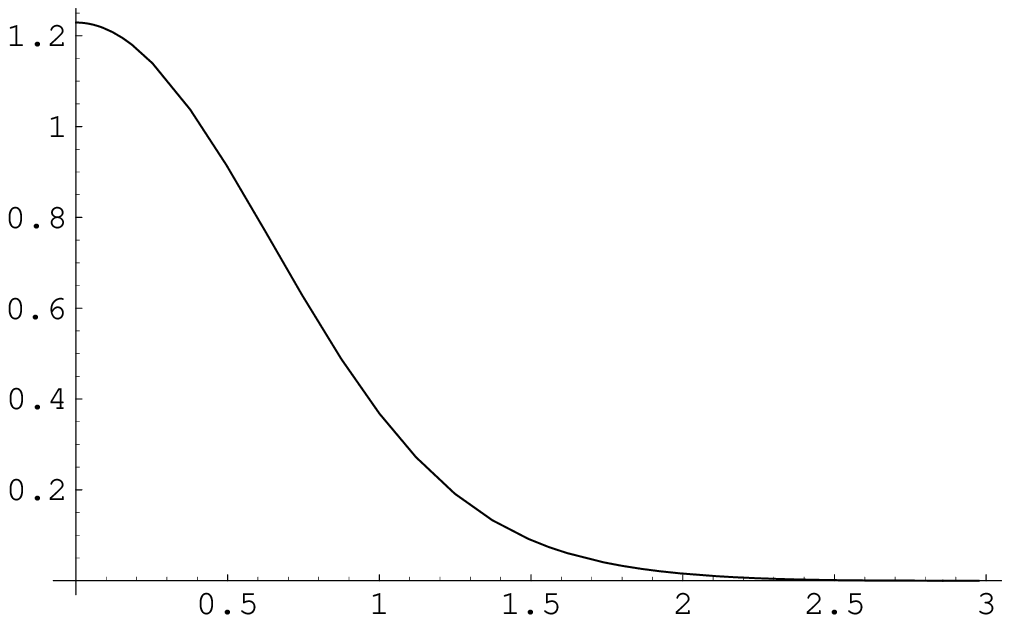}
{\hspace{-2.0cm}} {\hspace{-2.0cm}}
{\hspace{4.0cm} $\longrightarrow a \sqrt{u}$}\\
\rotatebox{90}{\hspace{-0.0cm} {$\Psi(a \sqrt{u}) \longrightarrow$}}
\includegraphics[scale=0.6]{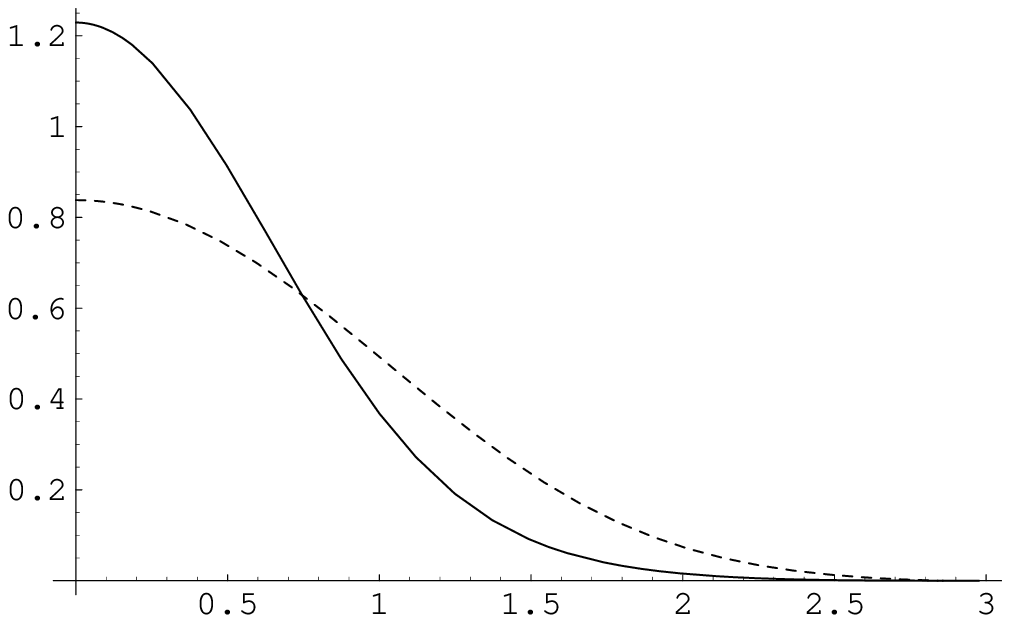}
 {\hspace{-2.0cm}} {\hspace{-2.0cm}} {\hspace{4.0cm}
$\longrightarrow a \sqrt{u}$}
 \caption{On the top we show the function $\Psi(a \sqrt{u})$ which enters the differential
 (with respect to the energy transfer u)
 event rate in dark matter searches. It has been obtained in the context of the Eddington
 approach  as discussed in the text.
For comparison we present at the bottom the same function
(continuous curve) together with that obtained using the M-B
distribution (dotted curve).}
 \label{psiu}
   \end{center}
  \end{figure}

 In Eq. (\ref{snew}) $\bar{K}$ is given by:
  \beq \bar{K}=\frac{\rho (0)}{100\mbox{ GeV}}
\frac{m}{m_p}~
              \sqrt{\langle v^2 \rangle }\simeq 160~10^{-4}~(pb)^{-1} y^{-1}\frac{\rho(0)}{0.3GeVcm^{-3}}
\frac{m}{1Kg}\frac{ \sqrt{\langle v^2 \rangle }}{280kms^{-1}}
\label{Kconst} \eeq and
\begin{equation}
c_{coh}(A, \mu_r(A))=\frac{100\mbox{ GeV}}{m_{\chi^0}}\left[
\frac{\mu_r(A)}{\mu_r(p)} \right]^2 A~t_{coh}(A) \label{ctm}
\end{equation}
\begin{equation}
c_{spin}(A, \mu_r(A))=\frac{100GeV}{m_{\chi^0}}\left[
\frac{\mu_r(A)}{\mu_r(p)} \right]^2 \frac{t_{spin}(A)}{A}
\label{ctm1}
\end{equation}
The parameters $c_{coh}(A,\mu_r(A))$, $c_{spin}(A,\mu_r(A))$,
which give the relative merit
 for the coherent and the spin contributions in the case of a nuclear
target compared to those of the proton,  have already been
tabulated \cite{JDV05}
 for energy cutoff $Q_{min}=0$ and $10$ keV.\\
Via  Eq. (\ref{snew}) we can  extract the nucleon cross section
from the data.
 The most interesting quantity is $t_{coh}(A)$. It is defined as:
 \beq
 t_{coh}=\int_{u_{min}}^{u_{max}} \frac{dt_{coh}}{du} du
 \label{tcoh}
 \eeq
 u is the energy transfer to the nucleus
 (in dimensionless units, see below)
$$u_{min}\Leftrightarrow \mbox{detector threshold}$$
$$u_{max}\Leftrightarrow \mbox{maximum WIMP velocity}$$
 \beq
 \frac{dt_{coh}}{du}= \sqrt{\frac{2}{3}} \frac{\upsilon_{par}}{\upsilon_0}T(u)~,~T(u)=a^2|F(u)|^2
 \Psi(a\sqrt{u}) \label{T.1} \eeq
 for the coherent mode and
  \beq
 t_{spin}=\int_{u_{min}}^{u_{max}} \frac{dt_{spin}}{du} du
 \label{tspin}
 \eeq
 \beq
 \frac{dt_{spin}}{du}= \sqrt{\frac{2}{3}} \frac{ \upsilon_{par}}{\upsilon_0}T(u)~,~T(u)=a^2F_{11}u)
\Psi(a\sqrt{u}) \label{T.2} \eeq
where $v_{par}=v_m$ in the present approach and $v_{par}=v_0$  in
the case of the M-B distribution and
The nucleon cross sections, which carry the dependence on the
particle model  parameters, are the most important ones, but they
 are not of interest in our present calculation. One such parameter is, of course, the WIMP mass.
 In the above expressions $F(u)$ is the form factor, entering the
coherent scattering
 and $F_{11}(u)$ is the spin response function entering via the axial current.
The function $\Psi$ depends on the WIMP distribution velocity
employed and is a function of the energy $Q$ transferred to the
nucleus \beq u=\frac{Q}{Q_0}~~,~~Q_{0}=\frac{4 A m_p}{b^2}=
4.1\times 10^{4}~A^{-4/3}~KeV \label{u.1} \eeq
 where $A$ is the nuclear mass number and the dimensionless parameter $a$ is given by:
\beq a=[\sqrt{2}\mu_r b \upsilon_{par}]^{-1}, \label{a.1} \eeq
where $\mu_r$ is the reduced mass of the WIMP-nucleus system and
$b$ is the
(harmonic oscillator) size parameter.\\
 The function, which is basic to us, $\Psi$, is given by
\beq
\Psi(a \sqrt{u})=~\int_{a \sqrt{u}}^{y_m}~ dy~\int_0^{\pi} ~ \sin{\theta} d\theta \int_0^{2\pi}~d\phi~y~
        f_v(y,\theta,\phi)
\label{psi.1}
\eeq
with
\beq
f_v(y,\theta,\phi) =f_v(\sqrt{y^2+2 y y_{sun} \cos{\theta}+y_{sun}^2})
\eeq
where $\theta$ is the  polar angle as measured from the direction of the sun's motion and $y_{sun}$ is the
sun's velocity in units of $v_m$. Note that since the argument of the function $f_v$ is constrained to be
less than $y_m$. If $y_m$ is small the allowed region in the $(y,\xi)$ space is very restricted.
The function $\Psi(a \sqrt{u})$ is plotted in Fig. \ref{psiu}.
 The function $\Psi$ for both distributions is shown in Fig. \ref{psiu}. We should stress that the
  relative differential  rate $\frac{1}{\bar{R}}\frac{dR}{du}$ can be obtained by combining the above
   results of $\Psi(a \sqrt{u}$ with  the nuclear form factor for the target of interest.
  The dependence on the  WIMP mass comes through the parameter $a$.
\section{Application in the case of the target $^{127}$I}
\begin{figure}[!ht]
 \begin{center}
\rotatebox{90}{\hspace{-0.0cm} {$F(u) \longrightarrow$}}
\includegraphics[scale=0.8]{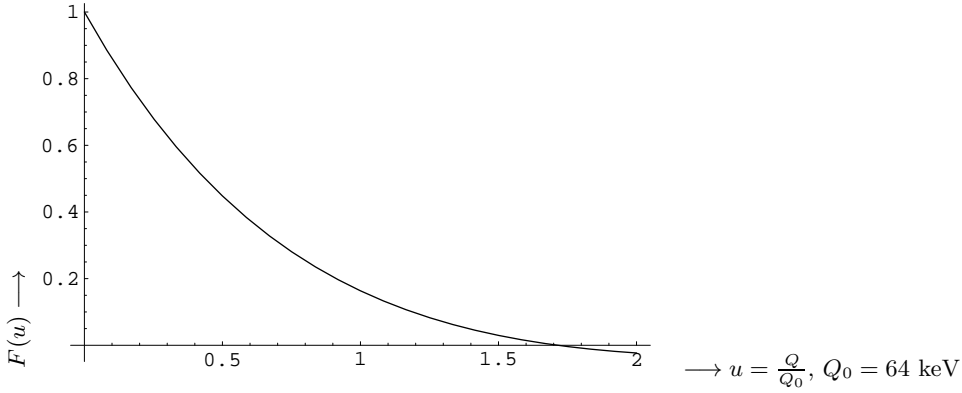}
{\hspace{-2.0cm}} {\hspace{-2.0cm}} {\hspace{4.0cm}
$\longrightarrow  u=\frac{Q}{Q_0}$, $Q_0=64$ keV}
 \caption{The form factor employed in our calculation. $u$ is the energy transfer to the nucleus in units of $Q_0$, i.e. $u=Q/Q_0$, with $Q_0=64$ keV.}
 \label{formf}
   \end{center}
  \end{figure}

\begin{figure}[!ht]
 \begin{center}
\rotatebox{90}{\hspace{-0.0cm} {$\frac{dt_{coh}}{du}
\longrightarrow$}}
\includegraphics[scale=0.5]{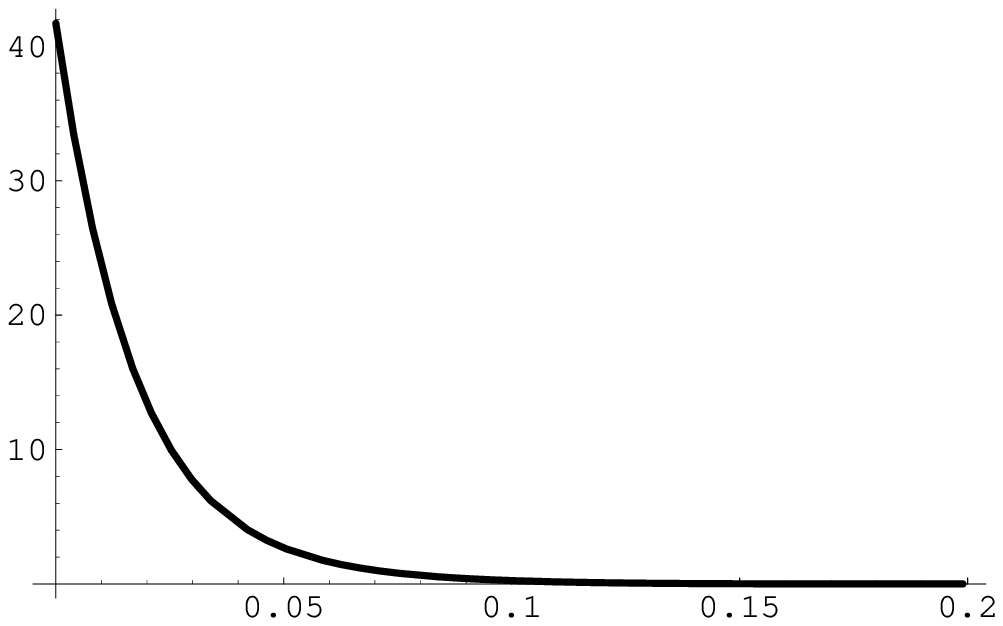}
\rotatebox{90}{\hspace{-0.0cm} {$\frac{dt_{coh}}{du}
\longrightarrow$}}
\includegraphics[scale=0.5]{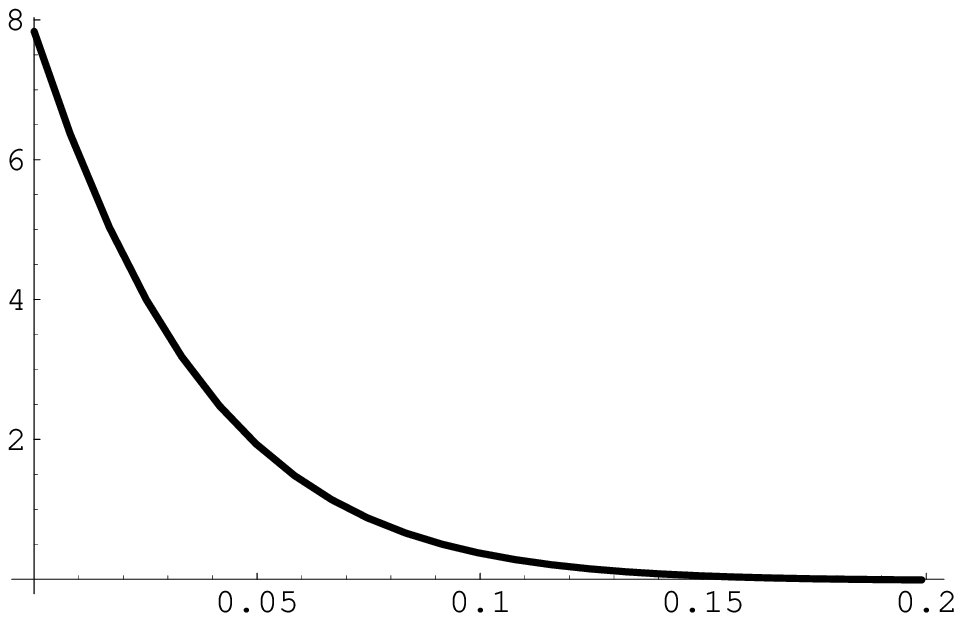}
{\hspace{-2.0cm}}
{\hspace{-2.0cm}}
{\hspace{4.0cm} $\longrightarrow  u=\frac{Q}{Q_0}$, $Q_0=64$ keV}
 \caption{On the left we show the quantity $ \frac{dt_{coh}}{du}$ for $m_{\chi}=10$ GeV
 in the case of the distribution obtained in this work. On the right we show
 the same quantity in the case of the M-B distribution. The
 expression for $t_{spin}$ is similar.
  Note that for such a small WIMP mass the differential rate drops very fast
   as a function of the energy transfer.}
 \label{mass1}
   \end{center}
  \end{figure}

\begin{figure}[!ht]
 \begin{center}
\rotatebox{90}{\hspace{-0.0cm} {$\frac{dt_{coh}}{du}
\longrightarrow$}}
\includegraphics[scale=0.5]{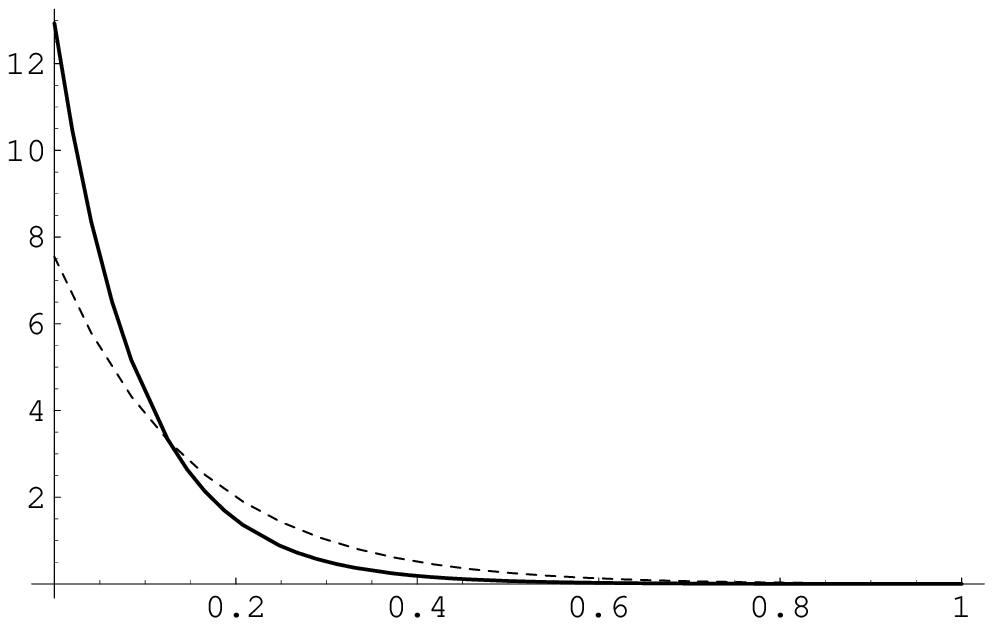}
\rotatebox{90}{\hspace{-0.0cm} {$\frac{dt_{coh}}{du}
\longrightarrow$}}
\includegraphics[scale=0.5]{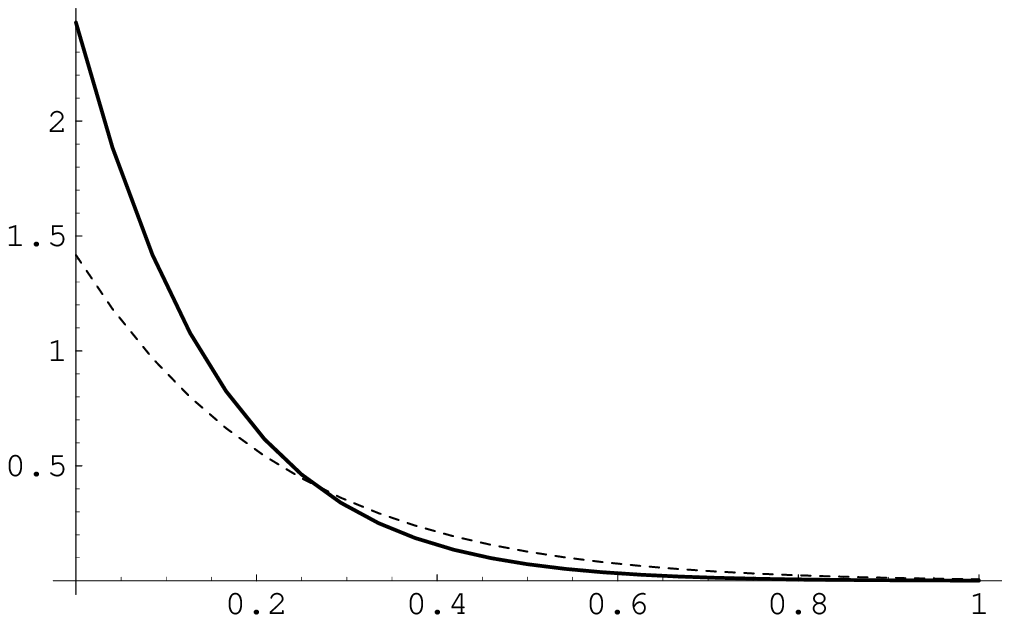}
{\hspace{-2.0cm}}
{\hspace{-2.0cm}}
{\hspace{4.0cm} $\longrightarrow  u=\frac{Q}{Q_0}$, $Q_0=64$ keV}
 \caption{The same as in Fig. \ref{mass1} for $m_{\chi}=30$ GeV (solid line) and $m_{\chi}=50$ GeV (dashed line).}
 \label{mass2}
   \end{center}
  \end{figure}

\begin{figure}[!ht]
 \begin{center}
\rotatebox{90}{\hspace{-0.0cm} {$\frac{dt_{coh}}{du}
\longrightarrow$}}
\includegraphics[scale=0.5]{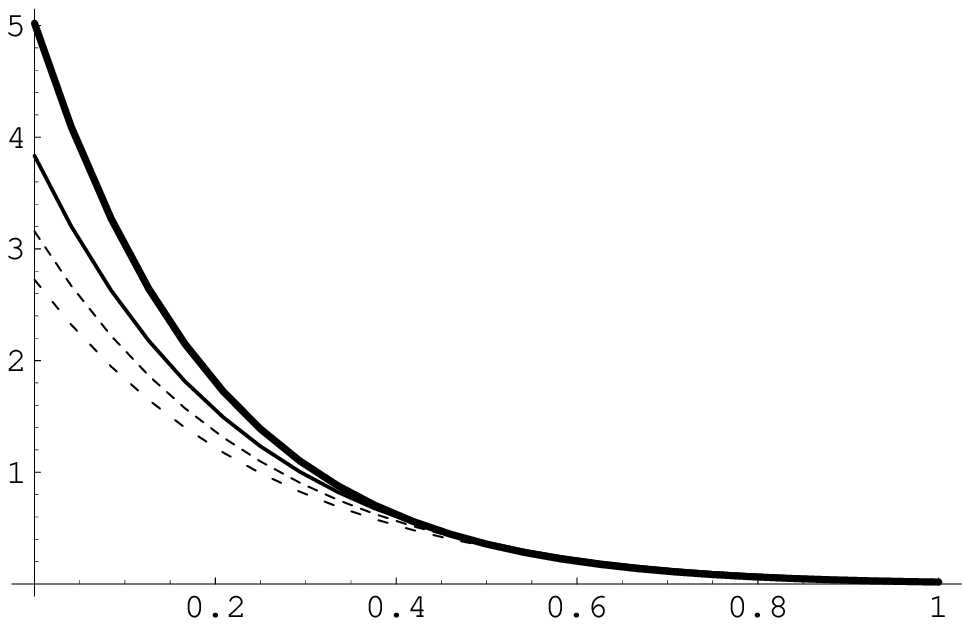}
\rotatebox{90}{\hspace{-0.0cm} {$\frac{dt_{coh}}{du}
\longrightarrow$}}
\includegraphics[scale=0.5]{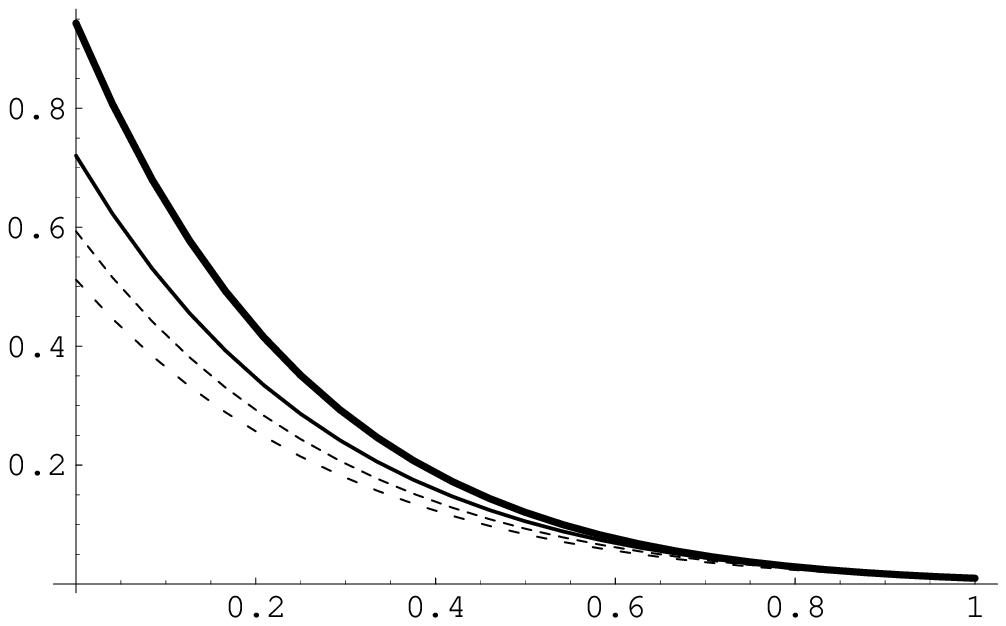}
\\ {\hspace{-2.0cm}} {\hspace{-2.0cm}} {\hspace{4.0cm}
$\longrightarrow  u=\frac{Q}{Q_0}$, $Q_0=64$ keV}
 \caption{The same as in Fig. \ref{mass1} for $m_{\chi}=75$ GeV (thick solid line), $m_{\chi}=100$ GeV (thin solid  line), $m_{\chi}=125$ GeV (short dashed line) and $m_{\chi}=150$ GeV  (long dashed line).}
 \label{mass3}
   \end{center}
  \end{figure}

\begin{figure}[!ht]
 \begin{center}
\rotatebox{90}{\hspace{-0.0cm} {$\frac{dt_{coh}}{du}
\longrightarrow$}}
\includegraphics[scale=0.5]{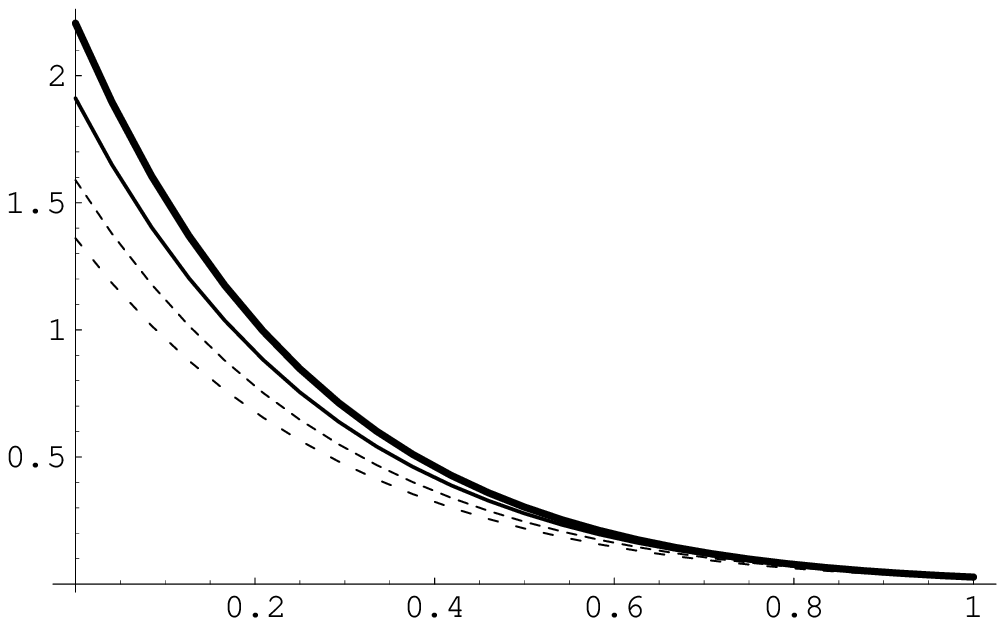}
\rotatebox{90}{\hspace{-0.0cm} {$\frac{dt_{coh}}{du}
\longrightarrow$}}
\includegraphics[scale=0.5]{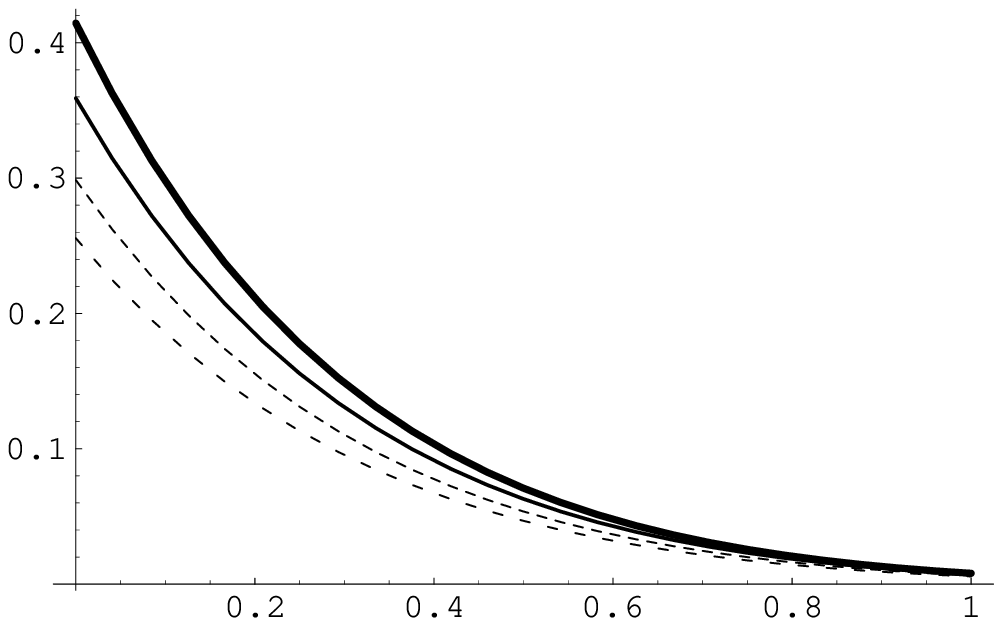}
{\hspace{-2.0cm}}
{\hspace{-2.0cm}}
{\hspace{4.0cm} $\longrightarrow  u=\frac{Q}{Q_0}$, $Q_0=64$ keV}
 \caption{The same as in Fig. \ref{mass3} for $m_{\chi}=200$ GeV (thick solid line), $m_{\chi}=250$ GeV (thin solid  line), $m_{\chi}=350$ GeV (short dashed line) and $m_{\chi}=500$ GeV  (long dashed line).}
 \label{mass4}
   \end{center}
  \end{figure}

\begin{figure}[!ht]
 \begin{center}
\rotatebox{90}{\hspace{-0.0cm} {$t_{coh} \longrightarrow$}}
\includegraphics[scale=0.5]{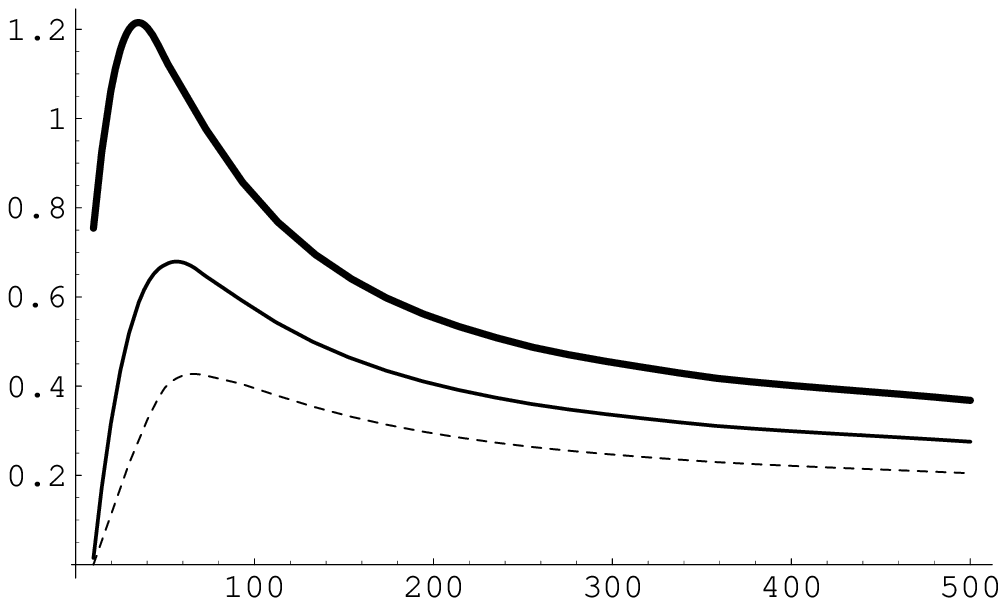}
\rotatebox{90}{\hspace{-0.0cm} {$t_{coh} \longrightarrow$}}
\includegraphics[scale=0.5]{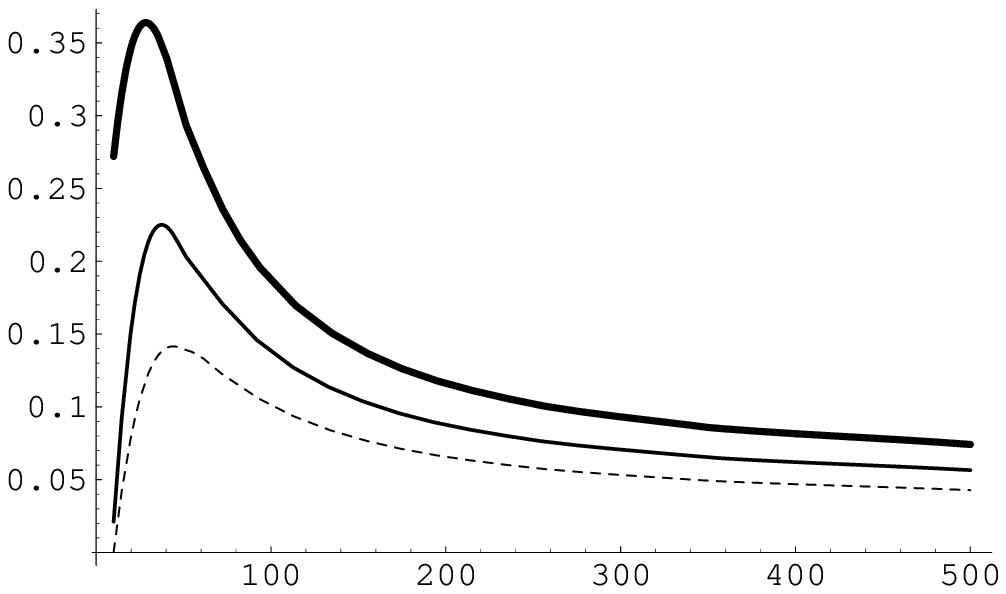}
{\hspace{-2.0cm}}
{\hspace{-2.0cm}}
{\hspace{4.0cm} $m_{\chi}\longrightarrow $ GeV}
 \caption{On the left we show the quantity $t_{coh}$ in the case of the present distribution as a function
  of the WIMP mass and threshold energy ($Q_{th}=0 \Longleftrightarrow$ thick solid curve,
   $Q_{th}=5$ keV $\Longleftrightarrow$ fine solid curve and $Q_{th}=10$ keV $\Longleftrightarrow$ dashed curve). On the right  we show the same quantity in the case of the M-B. It is clear that the rates decrease as the threshold energy increases. This is  is especially true for low LSP mass}
 \label{totrates}
   \end{center}
  \end{figure}

\begin{figure}[!ht]
 \begin{center}
\rotatebox{90}{\hspace{-0.0cm} {$t_{coh} \longrightarrow$}}
\includegraphics[scale=0.5]{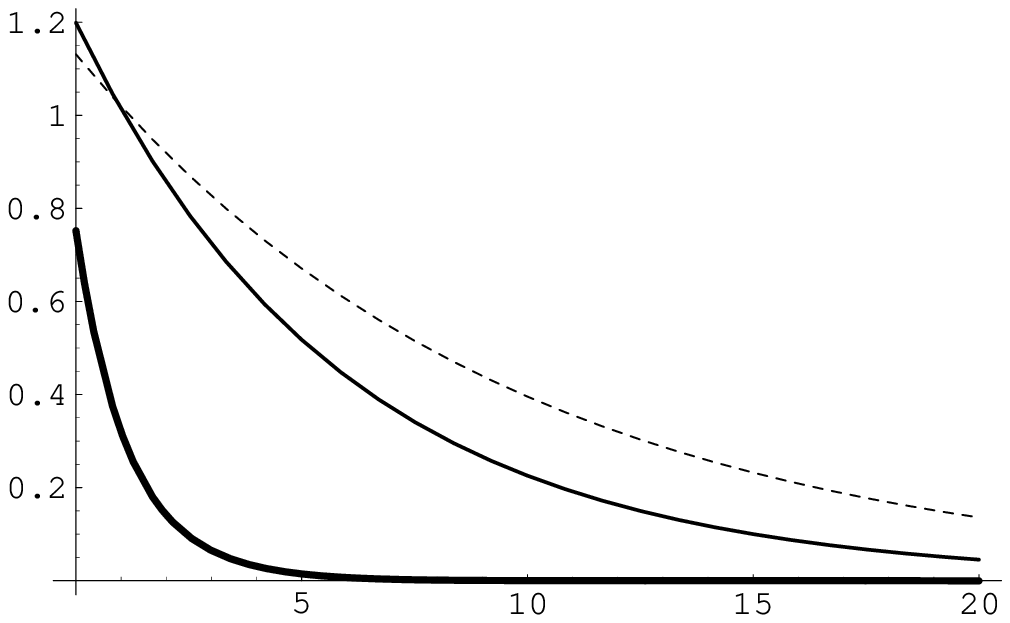}
\rotatebox{90}{\hspace{-0.0cm} {$t_{coh} \longrightarrow$}}
\includegraphics[scale=0.5]{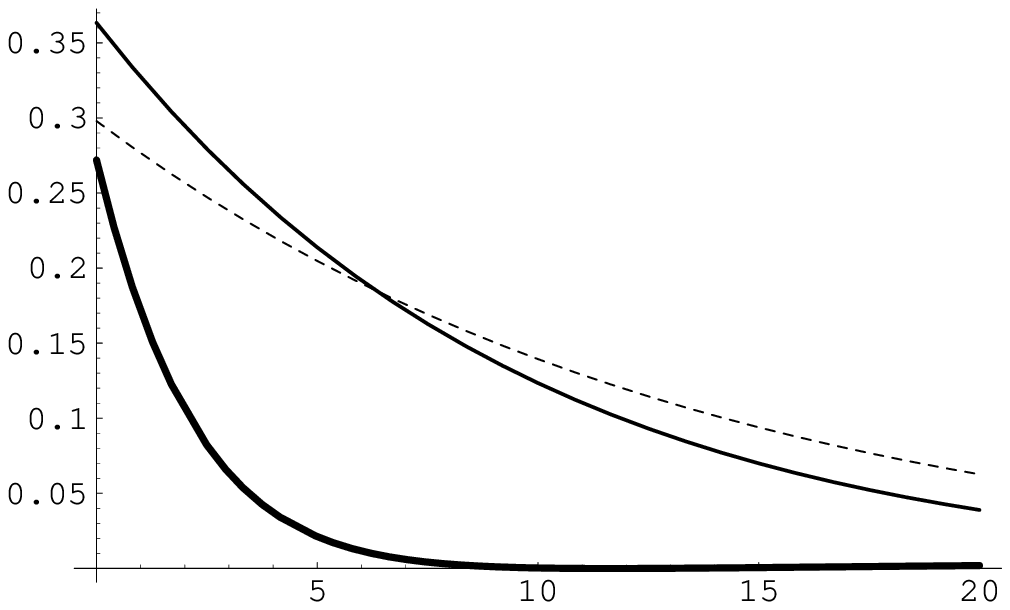}
{\hspace{-2.0cm}}
{\hspace{-2.0cm}}
{\hspace{4.0cm} $Q_{th} \longrightarrow$  keV}
 \caption{The relative rates $t_{coh}$ as a function of threshold energy for WIMP masses  $m_{\chi}=10$ (thick solid line), $m_{\chi}=30$ GeV (fine solid line) and $m_{\chi}=50$ GeV (dashed line). The results on the left correspond to the present distribution, while those on the right to the M-B distribution.}
 \label{ethres1}
   \end{center}
  \end{figure}

\begin{figure}[!ht]
 \begin{center}
\rotatebox{90}{\hspace{-0.0cm} {$t_{coh} \longrightarrow$}}
\includegraphics[scale=0.5]{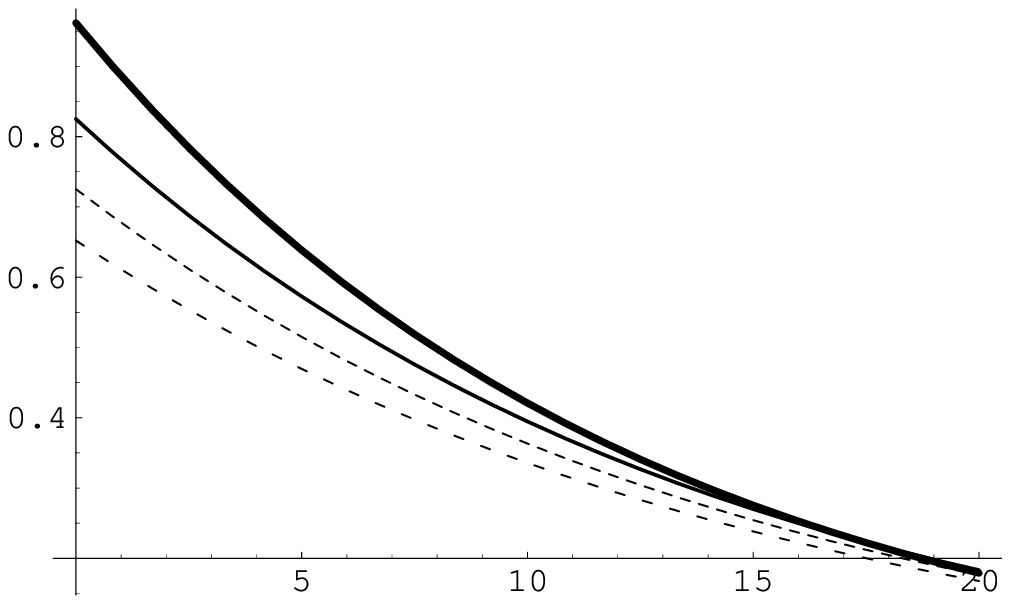}
\rotatebox{90}{\hspace{-0.0cm} {$t_{coh} \longrightarrow$}}
\includegraphics[scale=0.5]{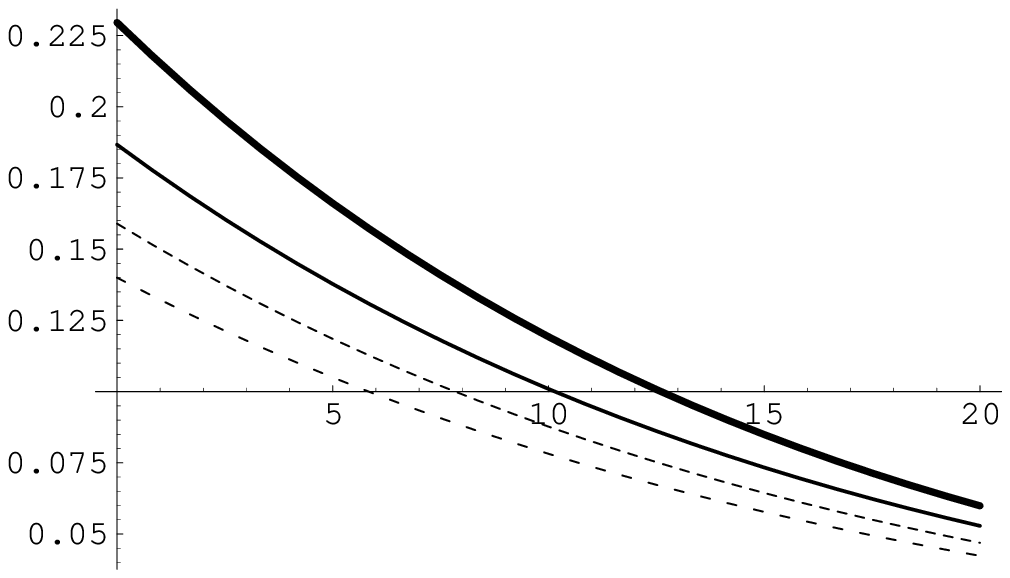}
{\hspace{-2.0cm}}
{\hspace{-2.0cm}}
{\hspace{4.0cm} $Q_{th} \longrightarrow$  keV}
 \caption{The same as in Fig. \ref{ethres1} for $m_{\chi}=75$ GeV (thick solid line), $m_{\chi}=100$
 GeV (thin solid  line), $m_{\chi}=125$ GeV (short dashed line) and $m_{\chi}=150$ GeV  (long dashed line).}
 \label{ethres2}
   \end{center}
  \end{figure}

\begin{figure}[!ht]
 \begin{center}
\rotatebox{90}{\hspace{-0.0cm} {$t_{coh} \longrightarrow$}}
\includegraphics[scale=0.5]{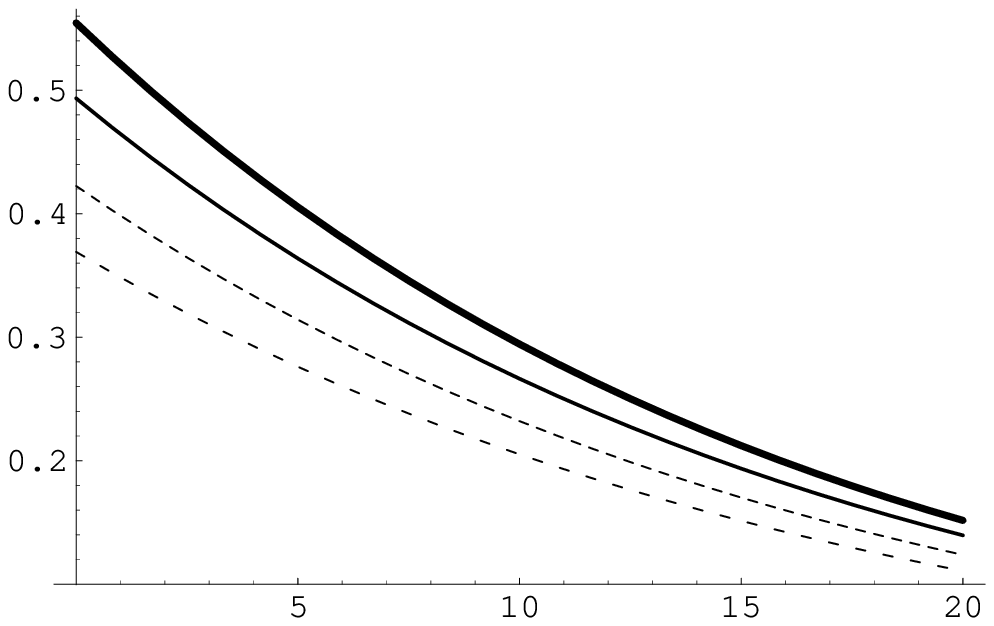}
\rotatebox{90}{\hspace{-0.0cm} {$t_{coh} \longrightarrow$}}
\includegraphics[scale=0.5]{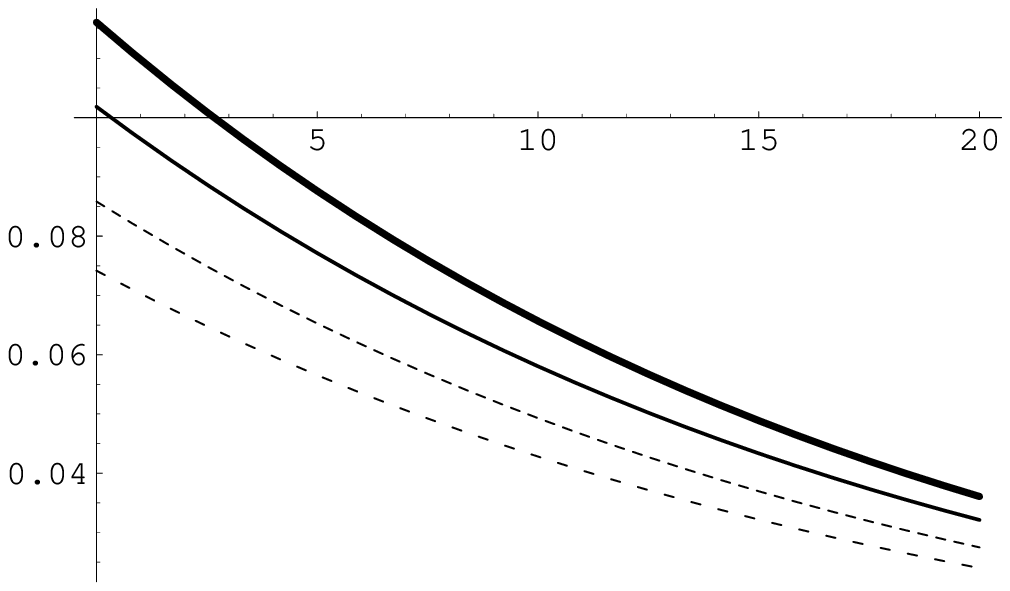}
{\hspace{-2.0cm}}
{\hspace{-2.0cm}}
{\hspace{4.0cm} $Q_{th} \longrightarrow$  keV}
 \caption{The same as in Fig. \ref{ethres2} for $m_{\chi}=200$ GeV (thick solid line), $m_{\chi}=250$
  GeV (thin solid  line), $m_{\chi}=350$ GeV (short dashed line) and $m_{\chi}=500$ GeV  (long dashed line).}
 \label{ethres3}
   \end{center}
  \end{figure}

In this section we are going to apply the above formalism in the
case of a popular target, $^{127}$I, which is an odd mass target
and can detect both the coherent and the spin modes of the
WIMP-nuclear interaction. We will include in our only the coherent
mode but we do not expect any real differences as far as the
quantity $\frac{1}{\bar{R}} \frac{dR}{du}$ is concerned.
 The nuclear form factor employed was obtained in the shell model description of the target and is shown in
 Fig. \ref{formf}.
  The quantity $ \frac{dt_{coh}}{du}$ obtained both for our distribution as well as for the familiar
  M-B distribution is shown in Figs \ref{mass1}-\ref{mass4} for various WIMP masses $m_{\chi}$.
  It is clear that the differential rate is a fast decreasing function of the energy transfer.
   This is particularly true for  low WIMP masses.\\
  Integrating the differential rate from zero to $u_{max}=y^2_{max}/a^2$, with $y_{max}=y_m$ ($y_{max=2.84}$)
  for the present (M-B) distributions respectively, we find the total rate $\frac{R}{\bar{R}}$ as a function
  of the WIMP mass. The results are shown in Fig. \ref{totrates}.
  As we have already mentioned the nucleon cross sections also depend on the LSP mass. So the absolute event
 rates, which include these cross sections, are expected to  drop even faster as a function of the LSP mass.
  In practice, however, the detectors have a low energy threshold. So only the event rates above an energy
transfer $Q_{th}$ can be detected. Thus we present the relative total rates $t_{coh}$ as a function
of $Q_{th}$ in figs \ref{ethres1}-\ref{ethres3}.
  From these plots we see that it is crucial for experiments to lower the threshold energy as much as possible.
   This is particularly true for small
   WIMP masses.
  \section{Modulation}
\begin{figure}[!ht]
 \begin{center}
\rotatebox{90}{\hspace{-0.0cm} {$H(a \sqrt{u}) \longrightarrow$}}
\includegraphics[scale=0.5]{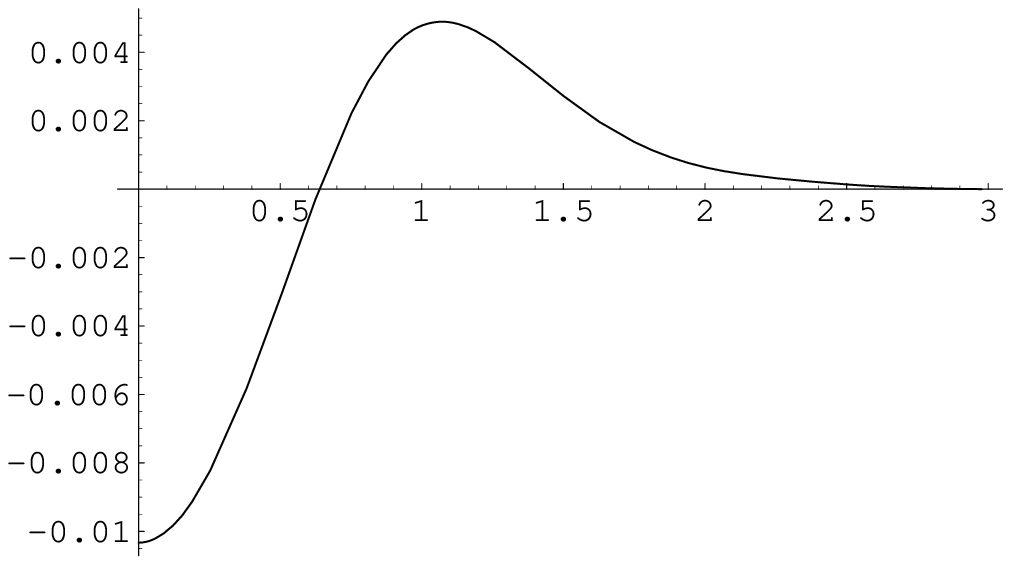}
\rotatebox{90}{\hspace{-0.0cm} {$H (a \sqrt{u}) \longrightarrow$}}
\includegraphics[scale=0.5]{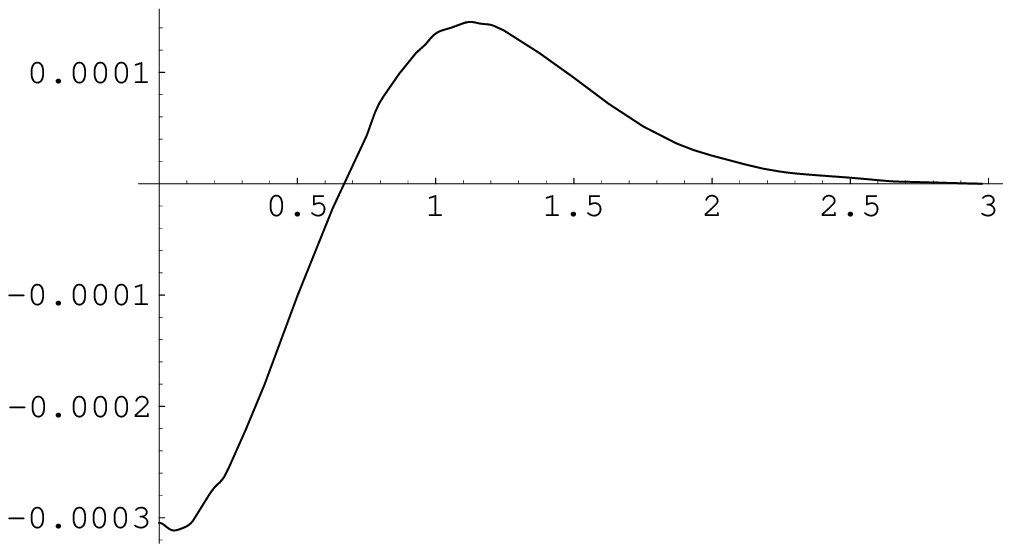}
{\hspace{-2.0cm}}
{\hspace{-2.0cm}}
{\hspace{4.0cm} $\longrightarrow a \sqrt{u}$}
 \caption{On the left we show the function $H(a \sqrt{u})$ governing the differential modulation amplitude
 as a function the energy transfer u. On the right we show the unimportant higher order effects
  $\sim \cos{2 \alpha}$.
  We note the sign change as the energy transfer increases.}
 \label{Hmod}
   \end{center}
  \end{figure}

\begin{figure}[!ht]
 \begin{center}
\rotatebox{90}{\hspace{-0.0cm} {$h \longrightarrow$}}
\includegraphics[scale=0.5]{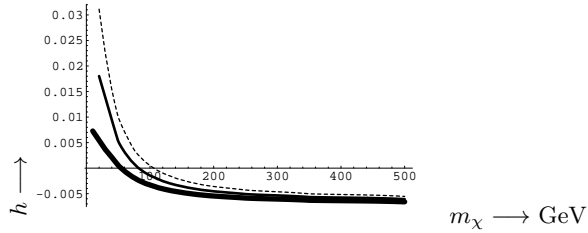}
{\hspace{-2.0cm}}
{\hspace{-2.0cm}}
{\hspace{4.0cm} $m_{\chi}\longrightarrow$ GeV}
 \caption{The modulation function $h$ is shown as a function of the WIMP mass for various threshold energies
 obtained with the present velocity distribution. In this figure:  $Q_{th}=0 \Longleftrightarrow$
  thick solid curve, $Q_{th}=5$ keV $\Longleftrightarrow$ fine solid curve and $Q_{th}=10$ keV
  $\Longleftrightarrow$ dashed curve.}
  \label{hmod}
   \end{center}
  \end{figure}

\begin{figure}
\begin{center}
\rotatebox{90}{\hspace{1.0cm} $h\rightarrow$}
\includegraphics[height=0.15\textheight]{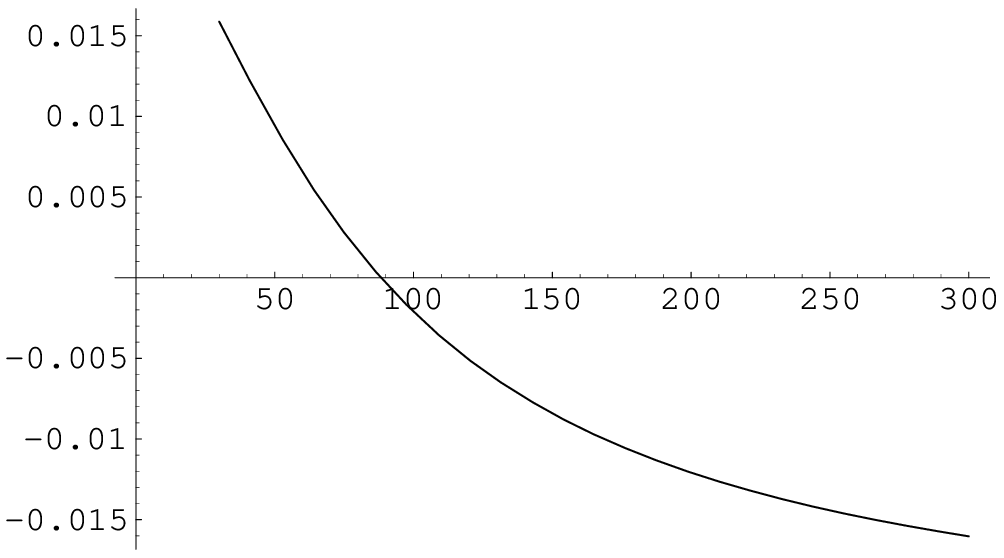}
\rotatebox{90}{\hspace{1.0cm} $h\rightarrow$}
\includegraphics[height=0.15\textheight]{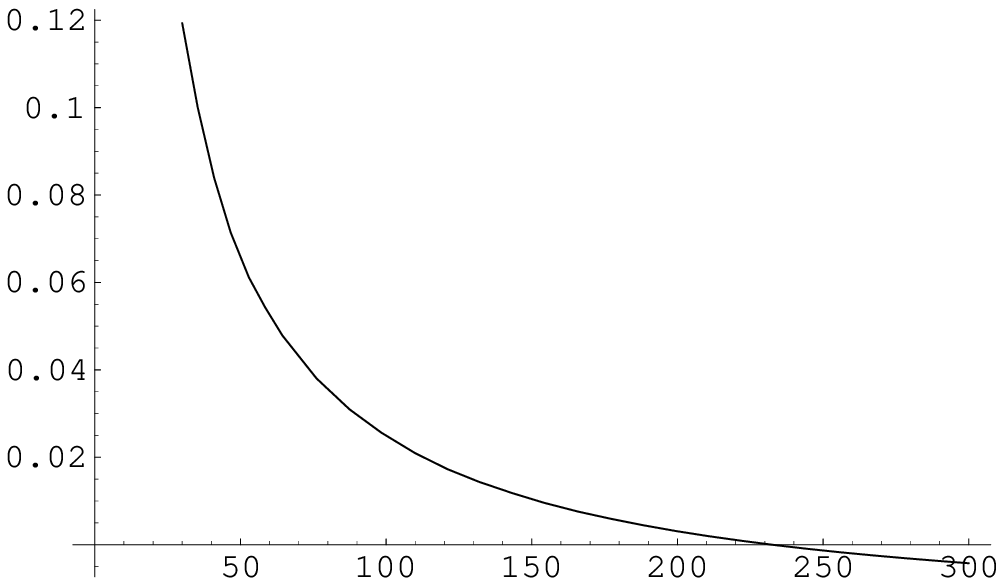}
 \hspace{0.cm}
 \hspace{0.5cm} $m_{\chi}\rightarrow$ ($GeV$)
\caption{ The modulation amplitude $h$ as a function of the WIMP
mass obtained with a M-B distribution in the case of $^{127}$I for
$Q_{min}=0$ on the left and $Q_{min}=10$ keV on the right.
  For the definitions see text.
\label{hgs} }
\end{center}
\end{figure}

  In the above discussion we did not take into account the motion of the Earth. The expected event rates,
  however, are very small due to the smallness of the nucleon cross sections not discussed in this work.
  So the experiments must fight against formidable backgrounds. Fortunately there are some signatures of
  the WIMP-nuclear interaction, which must be exploited. One such comes from the fact that the event rates
  depend on the relative velocity between the WIMP and the target. The most important velocity dependent
  contribution comes from the rotation of the Earth around the sun with a velocity$v_1=0.27~v_0$. It turns
  out that only the component of the Earth's velocity along the sun's direction of motion,
  $(v_1)_z=0.135 \cos{\alpha}$, is relevant ($\alpha$ is the phase of the Earth, $\alpha=0$ around
  June 3nd). We can thus apply the above formalism by
  $$y_{sun} \Rightarrow  y_{sun}(1+0.068 \cos{\alpha})$$
  Thus Eqs \ref{T.1} and \ref{T.2} become
  \beq
 T(u)\Longrightarrow a^2|F(u)|^2 \left[ \Psi(a\sqrt{u})
+H(a\sqrt{u}) \cos{\alpha} \right] \label{TH.1} \eeq for the
coherent mode and similarly for the spin
We will examine the modulation effect in the case of $^{127}$I. The behavior of the function $H(a\sqrt{u})$
 is exhibited in Fig. \ref{Hmod}. We notice the sign change of $H$ as the energy transfer changes.
  Integrating the differential rates we obtain  the total rates, which now now take the form:
\beq t\Longrightarrow t(1+h \cos{\alpha}) \label{heq} \eeq
of course, on the parameter $a$. Quite generally for a light
nuclear system and relatively heavy WIMP,
 the parameter $a$ is large, i.e. the lower part of the Fig
The exact behavior of the modulation of  event rates depends, of course, on the parameter $a$ and the nuclear
form factor. Quite generally for light nuclear targets, i.e. large $a$, the lower range of Fig. \ref{Hmod}
 does not enter and the modulation $h$ is not suppressed. On the other hand, for intermediate and heavy nuclei,
 the lower part tends to cancel the
effect of the upper part of Fig. \ref{Hmod}. This leads to a reduction of the modulation of the total rates and
 even in a change of sign of $h$, i.e. minimum in June and maximum in December.
 The behavior of the quantity $h$, in the case of the present velocity
distribution foe $^{127}I$, is exhibited in Fig. \ref{hmod}. We see that the modulation predicted by the present
distribution is much smaller than that
obtained with a M-B distribution  (see Fig. \ref{hgs}). This is true regardless of the  energy cut off. Similar results
have been obtained in a modified M.B. distribution \cite{TETRADIS} obtained in a model that couples gravity
to a scalar field. This way the characteristic velocity is not
$\upsilon_0$, the sun's rotational velocity, but $\prec \upsilon_d\succ \sqrt{2}\gg \upsilon_0$.
  We should emphasize that $h$ is the ratio of the modulated rate to the average rate. In
  other words the increase of $h$ as the threshold energy increases, results from the fact that the average
   rate (denominator) decreases much faster than the modulated rate (numerator) . So the increase in $h$,
   which is a good signal against background, is not cheap. It comes at the expense of the number of counts.
   \section{Conclusions}
   In the present paper we discussed the dependence of the direct dark matter event rates on the density and velocity distribution of WIMP's. This was done in a self consistent way by applying Eddington's approach.
    Even though at present only spherically symmetric matter distributions can be treated this way, one can draw
   the following conclusions:
   \begin{itemize}
   \item Simple dark matter distributions can adequately describe the rotational curves of both spherical and disc
galaxies. With the assumed form of the density its scale was determined by fitting the rotational velocity of the sun. The value of dark matter density extracted is in agreement with that hitherto employed in dark matter
   calculations, $\simeq 0.3$ GeV/cm$^3$.
   \item From the assumed density profile we obtained the distribution function. Evaluation of the distribution
   function in our vicinity yields the velocity distribution. This velocity distribution automatically vanishes for velocities larger than a velocity $\upsilon_m$. $\upsilon_m$ depends on the square root of the potential in
our vicinity.
   It is quite different from the rotational velocity, which depends on the square root of the derivative
   of the potential.
   \item The velocity distribution obtained is not very different from the standard M-B in the range of
   velocities of interest to dark matter searches.
   \item We focused our attention on the factor $t$ entering dark matter rates, which is independent of the
   nucleon cross section, i.e. it holds for all heavy WIMPs.
   The values obtained in our model are larger than those obtained using the M-B
   distribution. The relative magnitude depends on the WIMP mass.
   For small WIMP masses the obtained results  are 3-4 times larger than those of the M-B case.
   \item We also computed the modulation amplitudes $H(u)$ and $h$. We find that both are more suppressed
than those obtained with the M-B distribution. It seems that the
precise value of the modulation sensitively depend on the assumed
velocity distribution (the form of the function $H(u)$ is
independent of the assumed nuclear form factor). It appears,
though,  that the modulation obtained with the present velocity
distribution is perhaps too small to be of practical interest
experimentally.
\end{itemize}
 We did not consider in this work asymmetric velocity distributions. Another signature not discussed in this work is the asymmetry, with respect to the sun's direction of motion,
  expected in directional experiments \cite{DRIFT}. In such experiments, which detect not only the energy of the
   recoiling nucleus, but its direction of motion as well, the modulation is expected to be direction
   dependent.  It is expected to be quite large in some directions. Such effects in the context of
   the present approach are currently under study.
\section*{Acknowledgements}
This work was supported by the European Union under the contract
MRTN-CT-2004-503369 as well as the program PYTHAGORAS-1. The
latter
 is part of the
Operational Program for Education and Initial Vocational Training
of the Hellenic Ministry of Education under the 3rd Community
Support Framework and the European Social Fund.

\begin{thebibliography}{41}
\expandafter\ifx\csname natexlab\endcsname\relax\def\natexlab#1{#1}\fi
\expandafter\ifx\csname bibnamefont\endcsname\relax
  \def\bibnamefont#1{#1}\fi
\expandafter\ifx\csname bibfnamefont\endcsname\relax
  \def\bibfnamefont#1{#1}\fi
\expandafter\ifx\csname citenamefont\endcsname\relax
  \def\citenamefont#1{#1}\fi
\expandafter\ifx\csname url\endcsname\relax
  \def\url#1{\texttt{#1}}\fi
\expandafter\ifx\csname urlprefix\endcsname\relax\def\urlprefix{URL }\fi
\providecommand{\bibinfo}[2]{#2}
\providecommand{\eprint}[2][]{\url{#2}}

\bibitem[{MAX()}]{MAXIMA-1}
\bibinfo{note}{S. Hanary {\it et al}: {\it Astrophys. J.} {\bf 545}, L5
  (2000);\\ J.H.P Wu {\it et al}: {\it Phys. Rev. Lett.} {\bf 87}, 251303
  (2001);\\ M.G. Santos {\it et al}: {\it Phys. Rev. Lett.} {\bf 88}, 241302
  (2002)}.

\bibitem[{BOO()}]{BOOMERANG}
\bibinfo{note}{P. D. Mauskopf {\it et al}: {\it Astrophys. J.} {\bf 536}, L59
  (2002);\\ S. Mosi {\it et al}: {\it Prog. Nuc.Part. Phys.} {\bf 48}, 243
  (2002);\\ S. B. Ruhl {\it al}, astro-ph/0212229 and references therein.}

\bibitem[{DAS()}]{DASI}
\bibinfo{note}{N. W. Halverson {\it et al}: {Astrophys. J.} {\bf 568}, 38
  (2002)\\ L. S. Sievers {\it et al}: astro-ph/0205287 and references therein.}

\bibitem[{\citenamefont{{Smoot et al (COBE Collaboration)}}(1992)}]{COBE}
\bibinfo{author}{\bibfnamefont{G.~F.} \bibnamefont{{Smoot et al (COBE
  Collaboration)}}}, \bibinfo{journal}{Astrophys. J.}
  \textbf{\bibinfo{volume}{396}}, \bibinfo{pages}{L1} (\bibinfo{year}{1992}).

\bibitem[{\citenamefont{{Jaffe et al}}(2001)}]{flat01}
\bibinfo{author}{\bibfnamefont{A.~H.} \bibnamefont{{Jaffe et al}}},
  \bibinfo{journal}{Phys. Rev. Lett.} \textbf{\bibinfo{volume}{86}},
  \bibinfo{pages}{3475} (\bibinfo{year}{2001}).

\bibitem[{\citenamefont{{Spergel et al}}(2003)}]{SPERGEL}
\bibinfo{author}{\bibfnamefont{D.~N.} \bibnamefont{{Spergel et al}}},
  \bibinfo{journal}{Astrophys. J. Suppl.} \textbf{\bibinfo{volume}{148}},
  \bibinfo{pages}{175} (\bibinfo{year}{2003}).

\bibitem[{\citenamefont{{Bennett et al.}}(1995)}]{Benne}
\bibinfo{author}{\bibfnamefont{D.~P.} \bibnamefont{{Bennett et al.}}},
  \bibinfo{journal}{Phys. Rev. Lett.} \textbf{\bibinfo{volume}{74}},
  \bibinfo{pages}{2867} (\bibinfo{year}{1995}).

\bibitem[{\citenamefont{Goodman and Witten}(1985)}]{GOODWIT}
\bibinfo{author}{\bibfnamefont{M.~W.} \bibnamefont{Goodman}} \bibnamefont{and}
  \bibinfo{author}{\bibfnamefont{E.}~\bibnamefont{Witten}},
  \bibinfo{journal}{Phys. Rev. D} \textbf{\bibinfo{volume}{31}},
  \bibinfo{pages}{3059} (\bibinfo{year}{1985}).

\bibitem[{\citenamefont{Ellis and Roszkowski}(1992)}]{ELLROSZ}
\bibinfo{author}{\bibfnamefont{J.}~\bibnamefont{Ellis}} \bibnamefont{and}
  \bibinfo{author}{\bibfnamefont{L.}~\bibnamefont{Roszkowski}},
  \bibinfo{journal}{Phys. Lett. B} \textbf{\bibinfo{volume}{283}},
  \bibinfo{pages}{252} (\bibinfo{year}{1992}).

\bibitem[{ref()}]{ref2}
\bibinfo{note}{A. Bottino {\it et al.}, {\it Phys. Lett B} {\bf 402}, 113
  (1997).\\ R. Arnowitt. and P. Nath, {\it Phys. Rev. Lett.} {\bf 74}, 4592
  (1995); {\it Phys. Rev. D} {\bf 54}, 2374 (1996); hep-ph/9902237;\\ V. A.
  Bednyakov, H.V. Klapdor-Kleingrothaus and S.G. Kovalenko, {\it Phys. Lett. B}
  {\bf 329}, 5 (1994).}

\bibitem[{\citenamefont{Kosmas and Vergados}(1997)}]{KVprd}
\bibinfo{author}{\bibfnamefont{T.~S.} \bibnamefont{Kosmas}} \bibnamefont{and}
  \bibinfo{author}{\bibfnamefont{J.~D.} \bibnamefont{Vergados}},
  \bibinfo{journal}{Phys. Rev. D} \textbf{\bibinfo{volume}{55}},
  \bibinfo{pages}{1752} (\bibinfo{year}{1997}).

\bibitem[{\citenamefont{Vergados}(1996)}]{JDV96}
\bibinfo{author}{\bibfnamefont{J.~D.} \bibnamefont{Vergados}},
  \bibinfo{journal}{J. of Phys. G} \textbf{\bibinfo{volume}{22}},
  \bibinfo{pages}{253} (\bibinfo{year}{1996}).

\bibitem[{Dre()}]{Dree}
\bibinfo{note}{M. Drees and M. M. Nojiri, {\it Phys. Rev. D} {\bf 48}, 3843
  (1993); {\it Phys. Rev. D} {\bf 47}, 4226 (1993).}

\bibitem[{Che()}]{Chen}
\bibinfo{note}{T. P. Cheng, {\it Phys. Rev. D} {\bf 38}, 2869 (1988); H-Y.
  Cheng, {\it Phys. Lett. B} {\bf 219}, 347 (1989).}

\bibitem[{JDV({\natexlab{a}})}]{JDV06}
\bibinfo{note}{J. D. Vergados, On The Direct Detection of Dark Matter-
  Exploring all the signatures of the neutralino-nucleus interaction,
  hep-ph/0601064.}

\bibitem[{JEL()}]{JELLIS93}
\bibinfo{note}{J. Ellis , M. Karliner, Spin Structure Functions, CERH-TH
  7072/93.}

\bibitem[{Res()}]{Ress}
\bibinfo{note}{M. T. Ressell {\it et al.}, {\it Phys. Rev. D} {\bf 48}, 5519
  (1993); M.T. Ressell and D. J. Dean, Phys. Rev. C {\bf 56}, 535 (1997).}

\bibitem[{\citenamefont{Divari et~al.}(2000)\citenamefont{Divari, Kosmas,
  Vergados, and Skouras}}]{DIVA00}
\bibinfo{author}{\bibfnamefont{P.~C.} \bibnamefont{Divari}},
  \bibinfo{author}{\bibfnamefont{T.~S.} \bibnamefont{Kosmas}},
  \bibinfo{author}{\bibfnamefont{J.~D.} \bibnamefont{Vergados}},
  \bibnamefont{and} \bibinfo{author}{\bibfnamefont{L.~D.}
  \bibnamefont{Skouras}}, \bibinfo{journal}{Phys. Rev. C}
  \textbf{\bibinfo{volume}{61}}, \bibinfo{pages}{054612}
  (\bibinfo{year}{2000}).

\bibitem[{Dru()}]{Druk}
\bibinfo{note}{A. K. Drukier, K. Freeze and D. N. Spergel, {\it Phys. Rev.} D,
  {\bf 33}, 3495 (1986);\\ J.I. Collar et al., {\it Phys. Lett} B {\bf 275},
  181 (1992).}

\bibitem[{\citenamefont{Vergados}(2000)}]{Verg00}
\bibinfo{author}{\bibfnamefont{J.~D.} \bibnamefont{Vergados}},
  \bibinfo{journal}{Phys. Rev. D} \textbf{\bibinfo{volume}{62}},
  \bibinfo{pages}{023519} (\bibinfo{year}{2000}).

\bibitem[{\citenamefont{Sikivie}(1999)}]{SIKIVI1}
\bibinfo{author}{\bibfnamefont{P.}~\bibnamefont{Sikivie}},
  \bibinfo{journal}{Phys. Rev. D} \textbf{\bibinfo{volume}{60}},
  \bibinfo{pages}{063501} (\bibinfo{year}{1999}).

\bibitem[{\citenamefont{Sikivie}(1998)}]{SIKIVI2}
\bibinfo{author}{\bibfnamefont{P.}~\bibnamefont{Sikivie}},
  \bibinfo{journal}{Phys. Lett. B} \textbf{\bibinfo{volume}{432}},
  \bibinfo{pages}{139} (\bibinfo{year}{1998}).

\bibitem[{\citenamefont{Vergados}(2001)}]{Verg01}
\bibinfo{author}{\bibfnamefont{J.~D.} \bibnamefont{Vergados}},
  \bibinfo{journal}{Phys. Rev. D} \textbf{\bibinfo{volume}{63}},
  \bibinfo{pages}{06351} (\bibinfo{year}{2001}).

\bibitem[{\citenamefont{Green}(2001)}]{Green}
\bibinfo{author}{\bibfnamefont{A.~M.} \bibnamefont{Green}},
  \bibinfo{journal}{Phys. Rev. D} \textbf{\bibinfo{volume}{63}},
  \bibinfo{pages}{103003} (\bibinfo{year}{2001}).

\bibitem[{\citenamefont{Gelmini and Gondolo}(2001)}]{Gelmini}
\bibinfo{author}{\bibfnamefont{G.}~\bibnamefont{Gelmini}} \bibnamefont{and}
  \bibinfo{author}{\bibfnamefont{P.}~\bibnamefont{Gondolo}},
  \bibinfo{journal}{Phys. Rev. D} \textbf{\bibinfo{volume}{64}},
  \bibinfo{pages}{123504} (\bibinfo{year}{2001}).

\bibitem[{\citenamefont{Copi et~al.}(1999)\citenamefont{Copi, Heo, and
  Krauss}}]{Krauss}
\bibinfo{author}{\bibfnamefont{C.}~\bibnamefont{Copi}},
  \bibinfo{author}{\bibfnamefont{J.}~\bibnamefont{Heo}}, \bibnamefont{and}
  \bibinfo{author}{\bibfnamefont{L.}~\bibnamefont{Krauss}},
  \bibinfo{journal}{Phys. Lett. B} \textbf{\bibinfo{volume}{461}},
  \bibinfo{pages}{43} (\bibinfo{year}{1999}).

\bibitem[{\citenamefont{Green}(2002)}]{GREEN02}
\bibinfo{author}{\bibfnamefont{A.~M.} \bibnamefont{Green}},
  \bibinfo{journal}{Phys. Rev. D} \textbf{\bibinfo{volume}{66}},
  \bibinfo{pages}{083003} (\bibinfo{year}{2002}).

\bibitem[{\citenamefont{Eddington}(1916)}]{EDDIN}
\bibinfo{author}{\bibfnamefont{A.~S.} \bibnamefont{Eddington}},
  \bibinfo{journal}{NRAS} \textbf{\bibinfo{volume}{76}}, \bibinfo{pages}{572}
  (\bibinfo{year}{1916}).

\bibitem[{\citenamefont{Merritt}(1985)}]{MERRITT}
\bibinfo{author}{\bibfnamefont{D.}~\bibnamefont{Merritt}}, \bibinfo{journal}{A
  J} \textbf{\bibinfo{volume}{90}}, \bibinfo{pages}{1027}
  (\bibinfo{year}{1985}).

\bibitem[{\citenamefont{Ullio and Kamionkowski}(2001)}]{ULLIO}
\bibinfo{author}{\bibfnamefont{P.}~\bibnamefont{Ullio}} \bibnamefont{and}
  \bibinfo{author}{\bibfnamefont{M.}~\bibnamefont{Kamionkowski}},
  \bibinfo{journal}{JHEP} \textbf{\bibinfo{volume}{0103}}, \bibinfo{pages}{049}
  (\bibinfo{year}{2001}).\\
  J.F. Navarro, C.S. Frenk and S.M. White, ApJ {\bf 462} (1996)
  563.

\bibitem[{\citenamefont{Owen and Vergados}(2003)}]{OWVER}
\bibinfo{author}{\bibfnamefont{D.}~\bibnamefont{Owen}} \bibnamefont{and}
  \bibinfo{author}{\bibfnamefont{J.~D.} \bibnamefont{Vergados}},
  \bibinfo{journal}{Astrophys. J.} \textbf{\bibinfo{volume}{589}},
  \bibinfo{pages}{17} (\bibinfo{year}{2003}), \bibinfo{note}{astro-ph/0203923}.

\bibitem[{\citenamefont{Jeans}(1915)}]{JEANS}
\bibinfo{author}{\bibfnamefont{J.}~\bibnamefont{Jeans}},
  \bibinfo{journal}{MNRAS} \textbf{\bibinfo{volume}{76}}, \bibinfo{pages}{70}
  (\bibinfo{year}{1915}).

\bibitem[{\citenamefont{Hansen et~al.}(2006)\citenamefont{Hansen, Moore, Zemp,
  and Stadel}}]{HANSEN06a}
\bibinfo{author}{\bibfnamefont{S.~H.} \bibnamefont{Hansen}},
  \bibinfo{author}{\bibfnamefont{B.}~\bibnamefont{Moore}},
  \bibinfo{author}{\bibfnamefont{M.}~\bibnamefont{Zemp}}, \bibnamefont{and}
  \bibinfo{author}{\bibfnamefont{J.}~\bibnamefont{Stadel}},
  \bibinfo{journal}{JCAP} \textbf{\bibinfo{volume}{0601}}, \bibinfo{pages}{014}
  (\bibinfo{year}{2006}), \bibinfo{note}{astro-ph/0505420}.

\bibitem[{\citenamefont{Hansen and Moore}(2006)}]{HANSEN06b}
\bibinfo{author}{\bibfnamefont{S.~H.} \bibnamefont{Hansen}} \bibnamefont{and}
  \bibinfo{author}{\bibfnamefont{B.}~\bibnamefont{Moore}},
  \bibinfo{journal}{New Astronomy} \textbf{\bibinfo{volume}{11}},
  \bibinfo{pages}{333} (\bibinfo{year}{2006}),
  \bibinfo{note}{;astro-ph/041473}.

\bibitem[{\citenamefont{Vergados}(1998)}]{Verg98}
\bibinfo{author}{\bibfnamefont{J.~D.} \bibnamefont{Vergados}},
  \bibinfo{journal}{Phys. Rev. D} \textbf{\bibinfo{volume}{58}},
  \bibinfo{pages}{103001} (\bibinfo{year}{1998}).

\bibitem[{\citenamefont{Vergados}(1999)}]{Verg99}
\bibinfo{author}{\bibfnamefont{J.~D.} \bibnamefont{Vergados}},
  \bibinfo{journal}{Phys. Rev. Lett.} \textbf{\bibinfo{volume}{83}},
  \bibinfo{pages}{3597} (\bibinfo{year}{1999}).

\bibitem[{\citenamefont{Plummer}(1911)}]{PLUMMER}
\bibinfo{author}{\bibfnamefont{H.~C.} \bibnamefont{Plummer}},
  \bibinfo{journal}{MNRAS} \textbf{\bibinfo{volume}{71}}, \bibinfo{pages}{460}
  (\bibinfo{year}{1911}).

\bibitem[{\citenamefont{von Ziepel}(1908)}]{ZEIPEL}
\bibinfo{author}{\bibfnamefont{M.~H.} \bibnamefont{von Ziepel}},
  \bibinfo{journal}{Ann. Obs. Paris} \textbf{\bibinfo{volume}{25}},
  \bibinfo{pages}{F1} (\bibinfo{year}{1908}).
  \bibitem{LINPRING87}
  D.N.C. Lin and J.E. Pringle, ApJ. {\bf 320}, L87 (1987)

\bibitem[{JDV({\natexlab{b}})}]{JDV05}
\bibinfo{note}{J.D. Vergados, Direct SUSY Dark Matter Detection- Constraints on
  the Spin Cross Section, hep-ph/0512305.}

\bibitem[{\citenamefont{Brouzakis and Tetradis}(2006)}]{TETRADIS}
\bibinfo{author}{\bibfnamefont{N.}~\bibnamefont{Brouzakis}} \bibnamefont{and}
  \bibinfo{author}{\bibfnamefont{N.}~\bibnamefont{Tetradis}},
  \bibinfo{journal}{JCAP} \textbf{\bibinfo{volume}{0601}}, \bibinfo{pages}{004}
  (\bibinfo{year}{2006}).

\bibitem[{DRI()}]{DRIFT}
\bibinfo{note}{The NAIAD experiment B. Ahmed {\it et al}, Astropart. Phys. {\bf
  19} (2003) 691; hep-ex/0301039\\ B. Morgan, A. M. Green and N. J. C. Spooner,
  Phys. Rev. D {\bf 71} (2005) 103507; astro-ph/0408047.}

\end{thebibliography}

\end{document}